\documentclass[letterpaper,aps,prd,onecolumn,superscriptaddress,showpacs,preprintnumbers,nofootinbib]{revtex4}
\usepackage{graphicx}
\usepackage[intlimits,centertags]{amsmath}
\usepackage{amssymb,amsfonts}
\usepackage{gensymb}
\usepackage{mathrsfs}
\usepackage{hyperref}

\begin{document}

\title{Cosmogenic gamma-rays and the composition of cosmic rays}

\author{Markus~Ahlers} 
\affiliation{C.N.Yang Institute for Theoretical Physics, SUNY at Stony Brook, Stony Brook, NY 11794-3840, USA}

\author{Jordi Salvado}
\affiliation{Departament d'Estructura i Constituents de la Mat\`eria and
  Institut de Ciencies del Cosmos,
  Universitat de Barcelona,\\ 647 Diagonal, E-08028 Barcelona, Spain}

\begin{abstract}
We discuss the prospects of detecting the sources of ultra-high energy (UHE) cosmic ray (CR) nuclei via their emission of cosmogenic $\gamma$-rays in the GeV to TeV energy range. These $\gamma$-rays result from electromagnetic cascades initiated by high energy photons, electrons and positrons that are emitted by CRs during their propagation in the cosmic radiation background and are independent of the simultaneous emission of $\gamma$-rays in the vicinity of the source. The corresponding production power by UHE CR nuclei (with mass number $A$ and charge $Z$) is dominated by pion photo-production ($\propto A$) and Bethe-Heitler pair production ($\propto Z^2$). We show that the cosmogenic $\gamma$-ray signal from a single steady UHE CR source is typically more robust with respect to variations of the source composition and injection spectrum than the accompanying signal of cosmogenic neutrinos. We study the diffuse emission from the sum of extragalactic CR sources as well as the point source emission of the closest sources.
\end{abstract}

\pacs{13.85.Tp, 95.85.Pw, 98.70.Sa}

\preprint{YITP-SB-11-15}

\maketitle

\section{Introduction}\label{sec:introduction}

The origin and chemical composition of UHE CRs ($E>10^{18}$~eV) is a long-standing enigma in CR physics~\cite{Nagano:2000ve,Amsler:2008zzb}. The spectrum at these energies is expected to be dominated by extragalactic sources that have not yet been unambiguously identified. The average mass number of the composition can be inferred directly from atmospheric CR showers by measuring the elongation rate distribution and comparing this to simulations. Recent results of the Pierre Auger collaboration~\cite{Abraham:2010mj} indicate a transition of UHE CRs within the energy range $10^{18}$~eV to $4\times10^{19}$~eV from a light spectrum (consistent with protons) towards a heavier composition~\cite{Abraham:2010yv}. In contrast, the HiRes collaboration~\cite{Abbasi:2007sv} finds a mass composition compatible with that of a proton-dominated spectrum~\cite{Abbasi:2009nf}. 

Indirect evidence for the CR composition can come from various features seen in the spectrum. The ``ankle'' at about $3\times10^{18}$~eV seems to be a natural candidate for the transition between galactic and extragalactic CRs~\cite{Linsley:1963bk,Hill:1983mk,Wibig:2004ye}, but a lower energy crossover at the ``second knee'' at about $5 \times10^{17}$~eV has also been advocated for proton-dominated spectra~\cite{Berezinsky:2002nc,Fodor:2003ph}. Proton-dominance beyond the ankle is expected to be limited beyond the {\it Greisen-Zatspin-Kuz'min} (GZK) cutoff~\cite{Greisen:1966jv,Zatsepin:1966jv} due to resonant pion photo-production in the cosmic microwave background (CMB). Intriguingly, a suppression of the CR spectrum at the expected energy of about $5\times10^{19}$~eV has been observed at a statistically significant level~\cite{Abbasi:2007sv,Abraham:2008ru} and is consistent with a proton dominance at these energies. However, a similar feature could also originate from nuclear photo-disintegration of UHE CR nuclei in the cosmic radiation background (CRB), or from an {\it in situ} energy cut-off of the injection spectrum. 

In light of these yet inconclusive and even controversial experimental findings it is important to consider alternative cosmic messengers associated with the production and propagation of UHE CR nuclei. The sources of UHE CRs are also expected to emit high energy radiation in the form of neutrinos and $\gamma$-rays due to their interaction with ambient matter, radiation or magnetic fields inside or in the vicinity of the acceleration region. The strength and spectral energy distribution of this emission depends, however, not only on the uncertain chemical composition and injection spectrum of UHE CRs but also on the source environment. This dependence on source parameters is absent for the fluxes of cosmogenic neutrinos and $\gamma$-rays that are produced by CR interactions with the CRB and intergalactic magnetic fields (IGMFs); these contributions can be directly related to the observed spectrum of UHE CRs.

The dominant interactions of UHE CR nuclei (atomic number $A$ and charge $Z$) with the CRB are nuclear photo-disintegation~\cite{Stecker:1969fw,Puget:1976nz,Stecker:1998ib}, pion photo-production~\cite{Stecker:1978ah} and Bethe-Heitler (BH) pair production~\cite{Blumenthal:1970nn}. Photo-disintegration in the CMB leads to a fast break-up of heavy nuclei at energies of about $A\times 3\times10^{19}$~eV with an interaction length of the order of a few Mpc. The direct contribution of photo-disintegration to cosmogenic neutrinos and $\gamma$-rays is negligible~\cite{Anchordoqui:2006pd,Aharonian:2010te}. The dominant channel for cosmogenic neutrinos is the photo-production of charged pions in the CMB and their subsequent decay~\cite{Stecker:1978ah}. This process can be approximated by treating the nucleons of the nuclei as free protons and neutrons with energy $E/A$ resulting in an production threshold with CMB photons of about $A\times 5\times10^{19}$~eV. In the decay chain of the charged pion the three emerging neutrinos carry away about 3/4 of the total pion's energy. 

Photo-nucleon interactions produce roughly equal number of charged and neutral pions. Gamma-rays and $e^\pm$ produced via $\pi^0$ and $\pi^\pm$ decay, respectively, subsequently cascade on CRB photons via repeated $e^+ e^-$ pair production and inverse Compton scattering. The net result is a pile up of $\gamma$-rays at GeV-TeV energies, just below the threshold for further pair production on the diffuse optical background. Another contribution to this electromagnetic cascade comes from the BH production of $e^+e^-$ pairs. The corresponding energy loss length of CR nuclei is minimal at energies of $A\times2\times10^{19}$~eV and decreases with charge as $Z^2/A$. 

The energy loss rate $b\equiv{\rm d}E/{\rm d}t$ in electrons, positrons and $\gamma$-rays determines the bolometric flux of the cosmogenic GeV-TeV $\gamma$-ray signal and cosmogenic neutrinos. The energy loss via pion photo-production at CR energy $E$ can be approximated by the energy loss rate of nucleons $N$ as $b_{A,\gamma\pi}(E)\simeq A\, b_{N,\gamma\pi}(E/A)$. Since pion photo-production has a relatively high energy threshold the contribution of cosmogenic neutrinos and $\gamma$-rays from this channel typically depends on the maximal energy of the UHE CR emission. This can lead to a significant model dependence of the prediction. Bethe-Heitler $e^+e^-$ pairs are already produced at much lower energy. The interaction can be treated as a continuous energy loss with rate $b_{A,{\rm BH}}(E) = Z^2\,b_{p,{\rm BH}}(E/A)$, where $b_{p,{\rm BH}}$ is the BH energy loss of protons. The $Z^2$-dependence of this process and the low threshold makes this channel an important contributor to the $\gamma$-ray cascade.

We will study in the following the relative contributions of BH pair production and pion photo-production to the $\gamma$-ray signal for CR models involving heavy nuclei. In section~\ref{sec:propagation} we start with a brief review of the propagation of UHE CRs and the calculation of the spectra. The development of electromagnetic cascades and the corresponding diffuse $\gamma$-ray signal for CR models are discussed in section~\ref{sec:cascades}. We will focus here on two CR models, a proton-dominated spectrum with a low crossover and an iron-dominated CR model motivated by the chemical composition inferred by Auger. In section~\ref{sec:PSflux} we discuss the prospects to detect the closest sources of UHE CR nuclei as point sources of cosmogenic $\gamma$-rays. This possibility depends on the presence of an IGMFs during the development of the cascade as outlined in section~\ref{sec:magneticfields}. We finally conclude in section~\ref{sec:conclusion}.

\section{Propagation of Cosmic Ray Nuclei}\label{sec:propagation}

The main reactions of UHE CR nuclei during their cosmic evolution are nuclear photo-disintegration~\cite{Stecker:1969fw,Puget:1976nz,Stecker:1998ib}, pion photo-production~\cite{Stecker:1978ah} and BH pair production~\cite{Blumenthal:1970nn} on CRB photons. The angular-averaged differential rate of a transition between nuclei of type $i$ and $j$ with energy $E_i$ and $E_j$ is defined as
%%%%%%%%%%%%
\begin{equation}\label{eq:gamma}
\gamma_{i\to j}(z,E_i,E_j) \equiv
\frac{1}{2}\int\limits_{-1}^1\mathrm{d}\cos\theta\int\mathrm{d}
\epsilon\,(1-\beta
\cos\theta) n_\gamma(z,\epsilon)\frac{{\rm d}\sigma_{i\to j}}{{\rm d}E_j}(\epsilon')\,,
\end{equation}
%%%%%%%%%%%%
where $n_\gamma(z,\epsilon)$ is the energy distribution of isotropic background
photons at redshift $z$ and $\epsilon'=\epsilon\gamma(1-\beta
\cos\theta)$ the photon's energy in the rest frame of the nucleus with Lorentz boost $\gamma\simeq E_i/Am_p$. Besides the contribution of the CMB we use the cosmic infrared/optical background (CIB) from Ref.~\cite{Franceschini:2008tp}. Due to the cosmic evolution of the CRB density the interaction rates scale with redshift. Whereas the CMB evolution follows an adiabatic expansion, $n_\gamma(z,\epsilon) = (1+z)^2\,n_\gamma(0,\epsilon/(1+z))$, we assume that the CIB evolution follows the star formation rate as described in the appendices of Ref.~\cite{Ahlers:2009rf}. From Eq.~(\ref{eq:gamma}) we define the integrated interaction rate $\Gamma_{i\to j}(E_i) =\int{\rm d}E_j\gamma_{i\to j}(z,E_i,E_j)$ and the total interaction rate $\Gamma_{i}(E_i) = \sum_j\Gamma_{i\to j}(E_i)$.

Photo-disintegration of nuclei with large mass number $A$ is dominated by the giant dipole resonance (GDR) with main branches $A\to(A-1)+N$ and $A\to(A-2)+2N$ where $N$ indicates a proton or neutron~\cite{Stecker:1969fw,Puget:1976nz,Stecker:1998ib}. The GDR peak in the rest frame of the nucleus lies at about $20$~MeV for one-nucleon emission, corresponding to $E^A_{\rm GDR} \simeq A\times 2\times\epsilon^{-1}_{\rm meV}\times10^{10}$~GeV in the cosmic frame with photon energies $\epsilon = \epsilon_{\rm meV}$~meV. The secondary nuclei with atomic number $A-1$ and $A-2$ inherit the boost of the initial nucleon and lie close to the next GDR at $E^{A-1}_{\rm res}$ and $E^{A-2}_{\rm res}$, respectively. Hence, the initial flux of nuclei emitted from CR sources rapidly cascades down to lighter nuclei. This leads to a suppression of the flux above $E_{\rm res,A}/A$.

The most general evolution of primary and secondary nuclei in the CRB includes all possible photo-disintegration transitions between nuclides $(A,Z)$ competing with the decay of unstable nuclides. For simplicity, we follow the work of {\it Puget, Stecker \& Bredekamp} (PSB)~\cite{Puget:1976nz} and consider only one stable isotope per mass number $A$ in the decay chain of ${}^{56}$Fe. At energies below 10~MeV in the rest frame of the nucleus there exist typically a number of discrete excitation levels that can become significant for low mass nuclei. Above 30~MeV, where the photon wavelength becomes smaller than the size of the nucleus, the photon can interact via substructures of the nucleus. Out of these the interaction with quasi-deuterons is typically most dominant and forms a plateau of the cross section up to the pion production threshold at $\sim145$~MeV. We use the reaction code {\tt TALYS}~\cite{Goriely:2008zu} to evaluate the cross sections $\sigma_{A\to B}$ of the exclusive processes $(\gamma,N)$, $(\gamma,2N)$, $(\gamma,\alpha)$, $(\gamma,N\alpha)$ and $(\gamma,2\alpha)$ (N stands for p or n) for nuclides of the PSB-chain with $10\leq A\leq 56$. For the cross sections of light nuclei with mass numbers $A=2,3,4$ and $9$ we use the parametrization provided in Ref.~\cite{RachenTHESIS}. 

Resonant photo-nuclear interactions with CMB photons set in at energies of $5\times10^{19}$~eV per nucleon and becomes hence more important for low mass fragments of the photo-disintegration process. We follow the approach outlined in Ref.~\cite{RachenTHESIS} and approximate the total photo-nucleus interaction by the isospin averaged interaction rate of free nucleons as $\Gamma_{A,\gamma\pi}(z,E) \simeq A \Gamma_{N,\gamma\pi}(z,E/A)$~\cite{RachenTHESIS}. We also assume that the participating nucleon is removed from the nucleus and regard this as a contribution to one-nucleon losses. The nucleon-photon interaction rates can be determined using the Monte Carle package {\tt SOPHIA}~\cite{Mucke:1999yb}. We refer to the Appendices of Refs.~\cite{Ahlers:2009rf,Ahlers:2010ty} for further details of the calculation.

Another important energy loss of UHE CR nuclei is BH pair production via scattering off the CRB photons. Since this is a coherent process of the nucleons the energy loss scales as $Z^2$ where $Z$ is the charge number of the nucleus~\cite{Blumenthal:1970nn}. Bethe-Heitler pair production with differential rate $\gamma_{A,{\rm BH}}$ can be treated as a continuous energy loss process with a rate
\begin{equation}
b_{A,{\rm BH}}(z,E) \equiv \int {\rm d E'} (E-E') \gamma_{A,{\rm BH}}(z,E,E')\,,
\end{equation}
where $E$ and $E'$ are the energies of the nucleus before and after scattering, respectively. The energy loss of nuclei can be related to the loss of protons as $b_{A,{\rm BH}}(E) = Z^2b_{p,{\rm BH}}(E/A)$. The energy loss length $E/b(E)$ via BH pairs is typically much smaller than the interaction length of resonant photo-disintegration of heavy nuclei, which is of the order of $(4/A)$~Mpc. Hence the primary nuclei will be fully disintegrated within a few Mpc. The light secondaries with small charge per nucleus, like protons and helium, do not exhibit a $Z^2$-enhancement. However, photo-disintegration preserves the total number of nucleons. In particular, the full disintegration of an initial source spectrum of nuclei $Q_A(E)$ into protons is equivalent to a proton emission rate $Q_p(E) \simeq A^2Q_A(EA)$. We will see later that the energy loss via BH of actual models of UHE CR nuclei receives also some $Z^2$-enhancement via BH loss for nearby sources.

For the calculation of the diffuse spectra of UHE CR nuclei we assume that the cosmic source distribution is spatially homogeneous and isotropic. The comoving number density $Y_i = n_i/(1+z)^3$ of a nuclei of type $i$ is then governed by a set of Boltzmann equations of the form:
\begin{equation}\label{eq:boltzmann}
\dot Y_i = \partial_E(HEY_i) + \partial_E(b_iY_i)-\Gamma_{i}\,Y_i
+\sum_j\int{\rm d} E_j\,\gamma_{j\to i}Y_j+\mathcal{L}_i\,,
\end{equation}
together with the Friedman-Lema\^{\i}tre equations describing the cosmic expansion rate $H(z)$ as a function of the redshift $z$. This is given by \mbox{$H^2 (z) = H^2_0\,[\Omega_{\rm m}(1 + z)^3 + \Omega_{\Lambda}]$}, normalized to its value today of $H_0 \sim70$ km\,s$^{-1}$\,Mpc$^{-1}$, in the usual ``concordance model'' dominated by a cosmological constant with $\Omega_{\Lambda} \sim 0.7$ and a (cold) matter component, $\Omega_{\rm m} \sim 0.3$~\cite{Amsler:2008zzb}.    The time-dependence of the redshift can be expressed via ${\rm d}z = -{\rm d} t\,(1+z)H$. 

The term $\mathcal{L}_i$ in Eq.~(\ref{eq:boltzmann}) corresponds to the emission rate of CRs of type $i$ per comoving volume. For the CR injection spectrum of nuclei of mass number $A$ we use a power-law approximation with exponential cut-offs above $E_{\rm max}$ and below $E_{\rm min}$,
\begin{equation}\label{eq:QA}
Q_A(E) \propto E^{-\gamma}\exp(-E/E_{\rm max})\exp(-E_{\rm min}/E)\,.
\end{equation} 
To account for cosmic evolution of the spectral emission rate per comoving volume we introduce an energy-independent scaling of the form $\mathcal{L}_A(z,E) = \mathcal{H}(z)Q_A(E)$ where we use the approximation
\begin{equation}
\mathcal{H}(z) \equiv \mathcal{H}_0(1+z)^n\Theta(z-z_{\rm min})\Theta(z_{\rm max}-z)\,,
\label{eq:sourden2}
\end{equation} 
with $z_{\rm min} = 0$ and $z_{\rm max} = 2$ unless otherwise stated. Note that our ansatz for $\mathcal{L}_A(z,E)$ does not distinguish between the cosmic evolution of the CR source density and the evolution of the intrinsic emission rate. This distinction is not important for the calculation of the diffuse CR spectra, but plays a role in the prediction of neutrino and $\gamma$-ray point source fluxes associated with these CR sources. We will assume that the emission rate of CR sources is fixed and that their number density evolves with redshift. 

%%%%%%%%%%%%%%%%%
\begin{figure}[t]\centering
\includegraphics[width=0.9\linewidth]{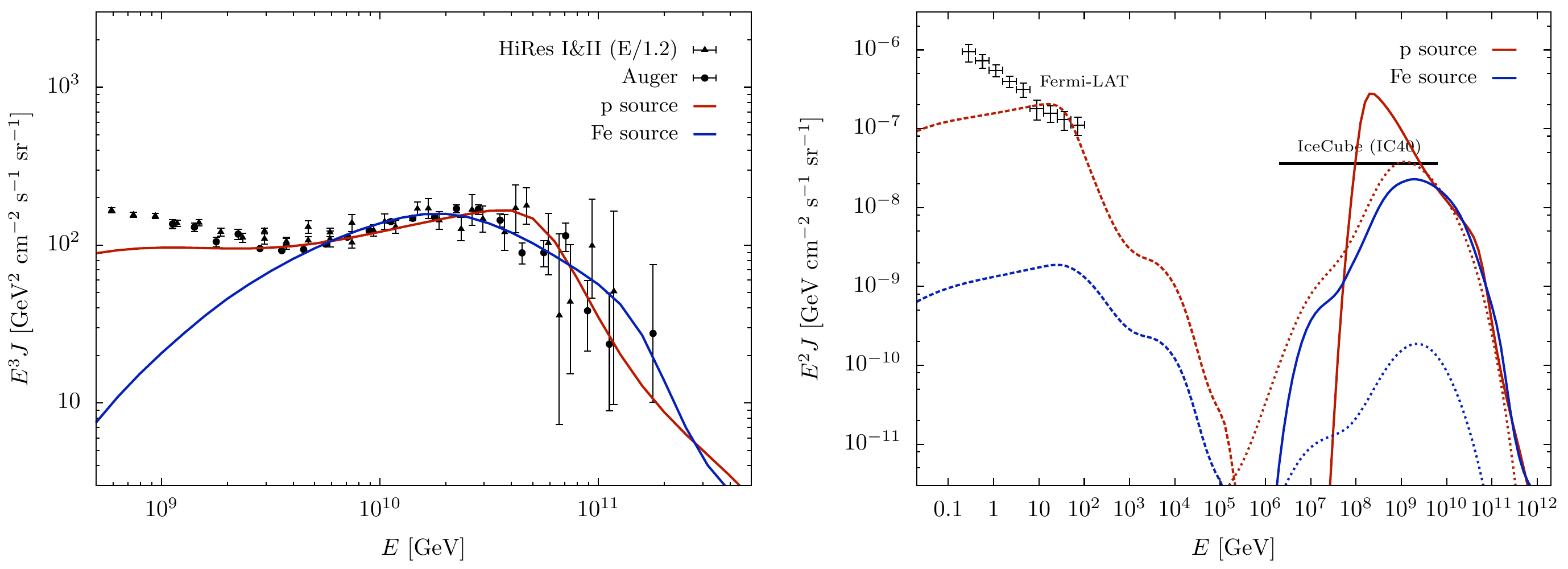}
\caption[]{{\bf Left panel:} Two models of extragalactic CRs assuming a homogenous distribution of protons (red line) and iron (blue line) between $z_{\rm min}=0$ and $z_{\rm max}=2$. For the proton sources we use an injection spectrum with $\gamma=2.3$, $E_{\rm min}=10^{18}$~eV, $E_{\rm max}=10^{20.5}$~eV and assume strong source evolution with $n=5$. The extragalactic iron sources assume an injection spectrum with $\gamma=2.3$, $E_{\rm min}=10^{18}$~eV, $E_{\rm max}=26\times10^{20.5}$~eV no evolution $n=0$. {\bf Right panel:} The corresponding spectra of cosmogenic $\gamma$-rays (dashed lines) and neutrinos (dotted line) for the two models. The diffuse $\gamma$-ray spectrum of the proton model is marginally consistent with the diffuse extragalactic spectrum inferred by Fermi-LAT~\cite{Abdo:2010nz} and the diffuse upper limit on cosmogenic neutrinos from the 40-string configuration (IC40) of IceCube~\cite{Abbasi:2011ji}. The cosmogenic $\gamma$-ray and neutrino spectra of the iron model are two orders of magnitude below the proton model predictions.}\label{fig1}
\end{figure}
%%%%%%%%%%%%%%%%%

In the following we are going to consider two models of extragalactic CR sources, that have been considered previously in fitting the UHE CR data~\cite{Berezinsky:2002nc,Allard:2008gj}. The first model consists of CR proton sources with a strong evolution ($n=5$) with a relatively low crossover below the ankle. For the injection spectrum we use the power index $\gamma=2.3$ and assume exponential cutoffs at $E_{\rm min}=10^{18}$~eV and $E_{\rm max}=10^{20.5}$~eV (see Eq.~(\ref{eq:QA})). The spectrum of protons after propagation through the CRB is shown as a red line in the left panel of Fig.~\ref{fig1}. The second model assumes a pure injection of iron with the same spectral index $\gamma=2.3$ but no evolution of the sources ($n=0$). We assume the same exponential cutoff at low energies as in the case of the proton model, $E_{\rm min}=10^{18}$~eV, and a high energy cutoff at $E_{\rm max}=26\times10^{20.5}$~eV, motivated by the rigidity dependence of the maximal energy of CR accelerators, $E_{\rm max} \propto Z$. The total spectrum of primary iron and secondary nuclei produced via photo-disintegration is shown as the blue line in the left panel of Fig.~\ref{fig1}. 

Both models reproduce the UHE CR data above the ankle reasonably well. The deficit below the ankle is assumed to be supplemented by a galactic contribution. Note that the crossover with the galactic component is higher for the all-iron model than for the all-proton model. The fit of the model spectra to the CR data sets the absolute normalization of the CR emission rate. This can be expressed as the required bolometric power density per CR source, which depends on the local density of source, ${\mathcal H}_0$. For both models we find a value of
\begin{equation}\label{eq:L19}
L \equiv \int {\rm d}E\,E\,Q(E)\simeq 10^{42}\left(\frac{\mathcal{H}_0}{10^{-5}\, {\rm Mpc}^{-3}}\right)^{-1}\,\, {\rm erg}\,  {\rm s}^{-1}\,.
\end{equation}

\section{Electromagnetic Cascades from Heavy Nuclei}\label{sec:cascades}

The evolution of cosmogenic electrons, positrons and $\gamma$-rays is governed by a set of Boltzmann equations analogous to Eqs.~(\ref{eq:boltzmann}). Electromagnetic interactions of photons and leptons with the CRB can happen on time-scales much shorter than their production rates~\cite{Lee:1996fp}. The driving processes of the electromagnetic cascade in the cosmic background photons are inverse Compton scattering (ICS) with CMB photons, $e^\pm+\gamma_{\rm bgr}\to e^\pm+\gamma$, and pair production (PP) with CMB and CIB radiation, $\gamma+\gamma_{\rm bgr}\to e^++e^-$~\cite{Blumenthal:1970nn,Blumenthal:1970gc}. In particular, the spectral energy distribution of multi-TeV $\gamma$-rays depends on the CIB background at low redshift. For our calculation we use the estimate of {\it Franceschini et al.}~\cite{Franceschini:2008tp}. We have little direct knowledge of the cosmic radio background. A theoretical estimate has been made by {\it Protheroe \& Biermann}~\cite{Protheroe:1996si} of the intensity down to kHz frequencies, based on the observed luminosity function and radio spectra of normal galaxies and radio galaxies although there are large uncertainties in the assumed evolution. The calculated values are about a factor of $\sim 2$ above the measurements and to ensure maximal energy transfer in the cascade we will adopt this estimate and assume the same redshift scaling as the CIB. However, the $\gamma$-ray cascade below TeV does not significantly depend on the exact value of this contribution. A summary of the CRB used in this calculation can be found in Fig.~A.6 of Ref.~\cite{Ahlers:2010fw}.

High energetic electrons and positrons may also lose energy via synchrotron radiation in the IGMF strength $B$, the strength of which is limited to be below $\sim 10^{-9}$G~\cite{Kronberg:1993vk,Beck:2008ty}. Recently, the absence of (resolvable) GeV emission from TeV $\gamma$-ray blazars has been used to infer a lower limit on the IGMF strength of the order of $10^{-15}$G ~\cite{Aharonian:1993vz,Neronov:2007zz,Neronov:1900zz,d'Avezac:2007sg,Tavecchio:2010mk}. The value of the IGMF has only little effect on the bolometric $\gamma$-ray flux in the GeV-TeV energy range relevant for our discussion~\cite{Ahlers:2010fw}, but a sufficiently low value is crucial for a discussion of point-source emission. In the calculation of the cascade spectrum we will assume a weak IGMF with strength of $10^{-15}$~G unless otherwise stated. Further processes contributing to the electromagnetic cascade are double pair production $\gamma+\gamma_{\rm bgr}\to e^++e^-+e^++e^-$ and triple pair production, $e^\pm+\gamma_{\rm bgr}\to e^\pm+e^++e^-$. These contribution have only a minor effect on the $\gamma$-ray flux at GeV-TeV and are neglected in the calculation for simplicity.

The observed diffuse extragalactic $\gamma$-ray flux is thought to be a superposition of various sources. Besides the cosmogenic $\gamma$-rays of UHE CR nuclei~\cite{Kalashev:2007sn} there are other candidates of truly diffusive processes associated with large-scale structure formation~\cite{Loeb:2000na} or models utilizing the decay and annihilation of dark matter~\cite{Abdo:2010dk}. Other contributors are unresolved extragalactic $\gamma$-ray sources like active galactic nuclei, starburst galaxies, or $\gamma$-ray bursts (see Ref.~\cite{Dermer:2007fg} for a recent review). A recent analysis of the diffuse extragalactic $\gamma$-ray background (EGRB) by Fermi-LAT~\cite{Abdo:2010nz} shows a $\gamma$-ray spectrum that is lower and softer than previous results of EGRET~\cite{Sreekumar:1997un}. It has been argued that the Fermi-LAT flux constraints all-proton models of UHE CRs extending down to energies of the ``second knee''~\cite{Berezinsky:2010xa}, though the systematics of UHE CR measurements is not sufficient to entirely exclude this model at a statistically significant level~\cite{Ahlers:2010fw}. 

The right panel of Fig.~\ref{fig1} shows the diffuse $\gamma$-ray spectra (dashed lines) from the all-proton model (red lines) and the all-iron with (blue lines). In both cases we normalize the CR spectra (solid lines) to the Auger data above the ankle. The contributions from both models differ by about two orders of magnitude which is in qualitative agreement with the previous study~\cite{Kalashev:2007sn}. Whereas the all-iron model has only a negligible contribution to the EGRB the proton model saturates the observed background at 10-100~GeV. (In fact, decreasing the lower cutoff $E_{\rm min}=10^{18}$~eV of the all-proton model would lead to an excess of the Fermi-LAT measurement.) For comparison, we also show cosmogenic neutrino flux (summed over flavors) of these models as dotted lines. The relative contributions from the two models differ by more than two orders of magnitude similar to the case of $\gamma$-rays. Note, that the cosmogenic neutrino flux from the all-proton model saturates a recent upper limit on the diffuse extragalactic neutrino flux from the 40-string sub-array (IC40) of IceCube~\cite{Abbasi:2011ji}. 

The strong model dependence of the diffuse fluxes is mainly due to the evolution of the sources as we will see in the following. The cascaded diffuse $\gamma$-ray flux peaks in the GeV-TeV region and has an almost universal shape here. Its normalization can be determined by the total energy loss rate into $\gamma$-rays, electrons and positrons during the propagation of UHE CR nuclei. We can define the comoving energy density at redshift $z$ as
\begin{equation}
\label{eq:omegacasz}
\omega_{\rm cas}(z) \equiv \int {\rm d} E\,E\left[Y_\gamma(z,E)+Y_{e^{-}}(z,E)+Y_{e^+}(z,E)\right]\,,
\end{equation}
which follows the evolution equation
\begin{equation}
\dot\omega_{\rm cas} + H\omega_{\rm cas} = \sum_A\int{\rm d} E\, b_A(z,E)Y_A(z,E)\,.
\end{equation}
The energy density (eV cm${}^{-3}$) of the electromagnetic background observed today is hence given by
\begin{equation}
\label{eq:omegacas}
\omega_{\rm cas}  = \sum_A\int{\rm
d}t\int{\rm d}E \,\frac{b_{A}(z,E)}{(1+z)}\,Y_{\rm A}(z,E)\,.
\end{equation}

The relative effect of cosmic evolution on the energy density of the cascade can be estimated in the following way. The UHE CR interactions with background photons are rapid compared to cosmic time-scales. The energy threshold of these processes scale with redshift $z$ as $E_{\rm th}/(1+z)$ where $E_{\rm th}$ is the (effective) threshold today. We can hence approximate the evolution of the energy density as
\begin{equation}
\dot\omega_{\rm cas} + H\omega_{\rm cas} \simeq \eta_{\rm cas}\mathcal{H}(z)\!\!\!\int_{E_{\rm th}/(1+z)}\!\!\!{\rm d} E\,E\,Q(E)\,,
\end{equation}
where $\eta_{\rm cas}$ denotes the energy fraction of the CR luminosity converted to the electromagnetic cascade. Assuming a power-law injection $Q(E)\propto E^{-\gamma}$ with sufficiently large cutoff $E_{\rm max}\gg E_{\rm th}$ we see that cosmic evolution enhances the diffuse $\gamma$-spectrum as
\begin{equation}\label{eq:evolfactor}
\omega_{\rm cas} \propto \int_{0}^{z_{\rm max}}\frac{{\rm d}z}{H(z)}(1+z)^{n+\gamma-3}\,.
\end{equation}
For the proton spectrum shown in Fig.~\ref{fig1} this corresponds to a relative factor of $\sim30$. 

An additional, yet smaller relative factor depend on the chemical composition. We start with the the energy loss via photo-nucleon interactions $b_{A,\pi\gamma}(E) \simeq Ab_{{\rm N},\pi\gamma}(E/A)$. This case is particularly simple to estimate: photo-disintegration losses conserve the total number of nucleons and we have the approximate relation $\sum_A A^2Y_A(EA) \simeq {\rm const}$. For the injection of a primary nucleus with mass number $A_0$ and power-law index $\gamma$ we expect the scaling $\omega_{\rm cas} \propto A_0^{2-\gamma}$ for a universal high energy cutoff per nucleon. 

Energy loss by BH pair production follows the scaling $b_{A,{\rm BH}}(E) \simeq Z^2b_{{\rm p},{\rm BH}}(E/A)$. In the absence of photo-disintegration and photo-nucleon interactions this would result in a simple scaling of the form $\omega_{\rm cas} \propto Z_0^2A_0^{1-\gamma}$. However, as we have already discussed in section~\ref{sec:propagation}, at those energies where pair-production is the dominant energy loss also nuclear photo-disintegration via the giant dipole resonance becomes important, shifting the average mass number and charge to lower values. On resonance, photo-disintegration has a typical inverse interaction rate of $(4/A)$~Mpc and hence the primary nuclei will be fully disintegrated within a few Mpc. In this case we can expect that the dominant contribution to BH loss comes from light secondary nuclei and we would obtain the scaling $\omega_{\rm cas} \propto A_0^{2-\gamma}$ analogous to the case of photo-nuclear losses. 

For nearby sources the competition between photo-nuclear processes and BH loss makes it difficult to predict the exact scaling of these quantities. For distant sources and in particular for the calculation of diffuse spectra we expect that the scaling is closer to $\omega_{\rm cas} \propto A_0^{2-\gamma}$. For the diffuse $\gamma$-spectra of the iron model shown in the right panel of Fig.~\ref{fig1} this corresponds to a relative factor $\sim0.3$ compared to the proton model. Together with the relative factor (\ref{eq:evolfactor}) from cosmic evolution and the difference in the normalization of the models ($Q_{{\rm Fe},{\rm GeV}}\simeq0.9 Q_{{\rm p},{\rm GeV}}$) this accounts for an overall factor of $\sim110$ in good agreement with the numerical value.

In summary, the contribution of UHE CRs to the EGRB depends strongly on the underlying CR model, in particular, the evolution of the sources. The CR spectrum at the highest energy is, however, dominated by local sources. The relative contribution from these point-sources does not depend on the evolution of the full population. As we will see in the following, the predicted $\gamma$-ray flux from these source is relatively robust against model variation of the emission spectrum.

%%%%%%%%%%%%%%%%%
\begin{figure}[t]\centering
\includegraphics[width=0.9\linewidth]{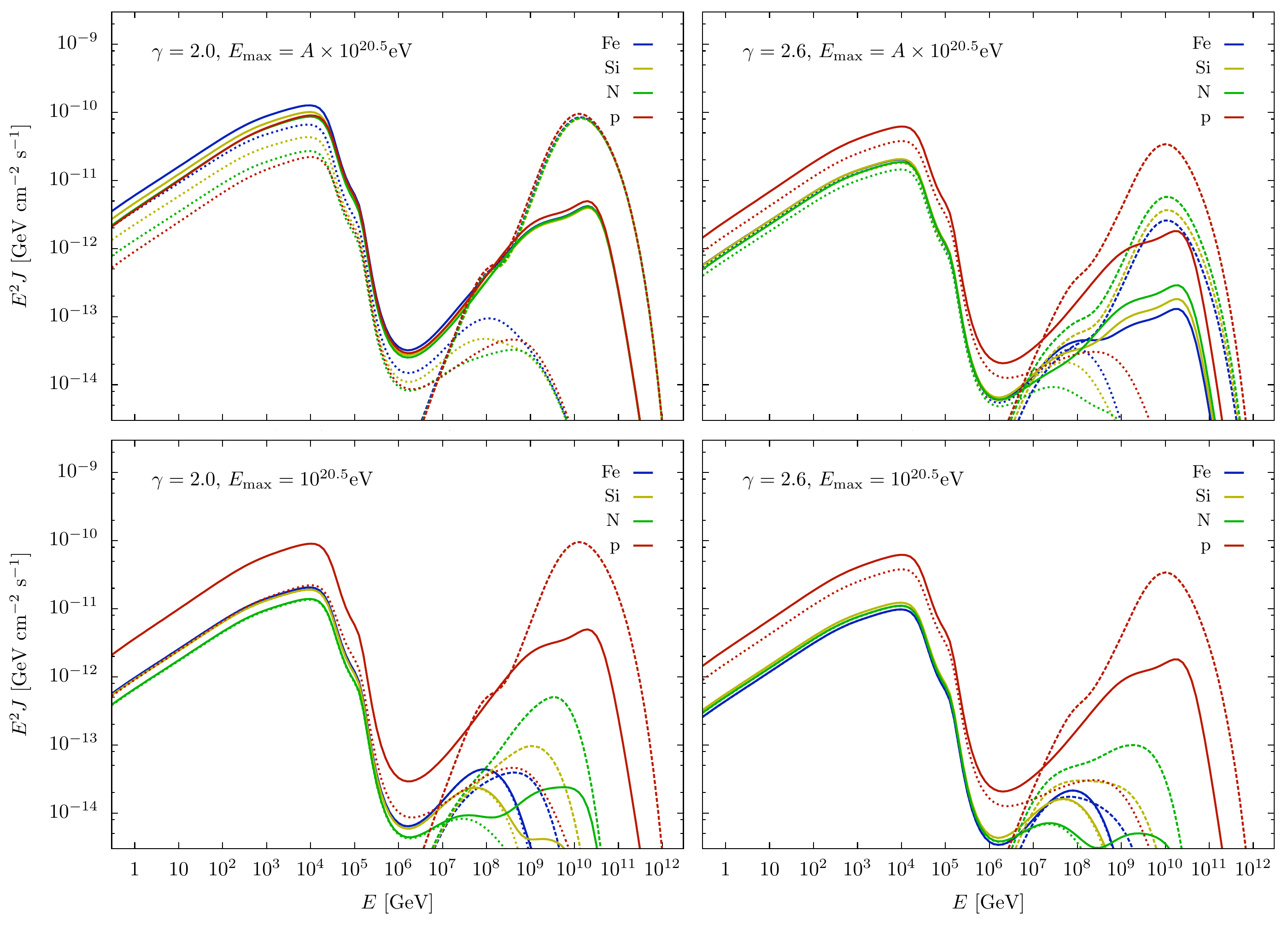}
\caption[]{The cosmogenic $\gamma$-ray and neutrino spectra of a CR point source of protons (red), nitrogen (green), silicon (yellow) and iron (blue) at a redshift $z=0.01$ ($\sim40$Mpc). We assume an injection spectrum of the form $Q_A(E)\propto E^{-\gamma}\exp(-E/E_{\rm max})$ with index $\gamma$ and cutoff $E_{\rm max}$ as indicated in the plots. For comparison, we chose the same normalization of the nuclei in each plot such that $L(E>10^{19}{\rm eV})=10^{42}$~erg/s (see Eq.~\ref{eq:L19}). The solid line shows the total flux of $\gamma$-rays and the dashed line the cosmogenic neutrino flux. The dotted lines show the contribution to $\gamma$-rays from BH loss alone omitting the contribution from photo-pion interactions.}\label{fig2}
\end{figure}
%%%%%%%%%%%%%%%%%

\section{Point-Source Flux}\label{sec:PSflux}

So far, we have only considered diffuse $\gamma$-ray fluxes from a spatially homogenous and isotropic distribution of CR sources. However, the discreteness of CR sources can lead to local $\gamma$-ray excesses, in particular for the closest sources. Under optimal circumstances, {\it i.e.}~sufficiently weak IGMFs, these excesses may even contribute as TeV $\gamma$-ray point-sources (PSs) in $\gamma$-ray observatories. This possibility has been previously studied for proton sources in Refs.~\cite{Ferrigno:2004am,Gabici:2005gd,Armengaud:2005cr,Kotera:2010xd} and has been revived recently in the context of unusually bright though distant TeV $\gamma$-ray sources~\cite{Essey:2009zg,Essey:2009ju,Essey:2010er}. Here we extend the discussion to the case of UHE CR nuclei and study the effect of IGMFs on the observability of the PS flux in detail.

In the absence of an IGMF the flux from a PS at redshift $z_\star$ with emission rate $Q_A(E)$ [GeV${}^{-1}$ s${}^{-1}$] as in Eq.~(\ref{eq:QA}) is equivalent to an integrated diffuse flux $4\pi J(E)$ from a homogenous distribution on a sphere at comoving distance $d_C(z) \equiv \int_0^z{\rm d}z'/H(z')$. The corresponding emission rate density is hence
\begin{equation}
\mathcal{L}^\star(z,E) = \frac{Q_A(E)}{4\pi d_C^2(z_\star)}H(z_\star)\delta(z-z_\star)\,.
\end{equation}
It is easy to check, that in the absence of interactions of the primary particle with the photon background the point source flux is then given as $J^\star(E) = 4\pi J(E) = Q_A((1+z_\star)E)/(4\pi d^2_C)$. In particular, this reproduces the familiar luminosity-distance relation $F=L/(4\pi d^2_L)$ with flux $F = \int {\rm d} E E J^\star$, luminosity $L = \int {\rm d} E E Q_0$ and luminosity distance $d_L = (1+z)d_C$.

For illustration, we show in Fig.~\ref{fig2} the $\gamma$-ray (solid lines) and neutrino (dashed lines) PS fluxes from a source at redshift $z_\star=0.01$ ($d_C(z_\star)\simeq40$~Mpc) emitting iron ($A=56$), silicon ($A=28$), nitrogen ($A=14$) or protons. The different plots show variations of the spectral index $\gamma$ of the injection spectrum and the exponential cutoff $E_{\rm max}$. In each plot the overall normalization of the proton emission rate $Q_p$ is chosen such that the source luminosity is $L_{p}(E>10^{19}\,{\rm eV})=10^{42}$~erg/s (see Eq.~\ref{eq:L19}) and we use the same normalization constant for the other nuclei. The top panels show the results of $\gamma=2.0$ (top left panel) and $\gamma=2.6$ (top right panel) for a cutoff proportional to the nucleon mass $E_{\rm max} = A\times10^{20.5}$~eV. This choice corresponds to a universal exponential cutoff for the energy per nucleon in each injection spectrum. Since the energy loss $b_{A,\gamma\pi}$ from photo-nucleon interactions depend on $E/A$ the spectra of cosmogenic neutrinos (dashed lines) have an almost universal shape as expected. Due to the universality of the neutrino spectra the relative normalization of their flux is in this case also given by Eq.~(\ref{eq:omegacas}) and scales as $A^{2-\gamma}$. 

The contribution of BH pair production to the $\gamma$-ray spectrum is shown separately in the plots as dotted lines. As discussed earlier, this contribution does in general not follow the $A^{2-\gamma}$ behavior of the cosmogenic neutrino fluxes. If the initial nucleus is not fully photo-disintegrated, the BH contribution will be closer to $Z^2A^{1-\gamma}$. Qualitatively, the variation of this contribution with spectral index $\gamma$ and initial mass number $A$ is smaller than for the case of photo-pion loss and stays within a factor $\sim5$. 

The robustness of this contribution becomes even more apparent in the case of a fixed maximal cutoff $E_{\rm max}=10^{20.5}$~eV for all nuclei which is shown in the lower panels of Fig.~\ref{fig2} for the same spectral indices $\gamma=2$ (bottom left panel) and $\gamma=2.6$ (bottom right panel). Since pion photo-production has a relatively high threshold of about $A\times5\times10^{19}$~eV their contribution becomes strongly suppressed as we go to heavier nuclei and hence lower maximal energy per nucleon. This is apparent from the drastic decrease of the cosmogenic neutrino flux (dotted lines). (Similarly, the GZK $\gamma$-ray flux~\cite{Gelmini:2007jy,Hooper:2010ze} at the upper end of the spectrum varies strongly with the maximal cutoff and compositon.) The $\gamma$-ray cascade, however, receives contributions from BH pair production at a lower energy threshold and this contribution is only mildly effected by the variation of the model parameters. Generally, the comparatively small variation of the BH contribution in the electromagnetic cascade makes the prediction of $\gamma$-rays from the sources of UHE CRs more robust than cosmogenic neutrinos. 

The fit of UHE CR models to the data fixes the average luminosity density of CR sources. For the prediction of the average luminosity per source and hence the cosmogenic $\gamma$-ray flux from the closest CR source we have to fix the local source density $\mathcal{H}_0$ [cm${}^{-3}$] introduced in Eq.~(\ref{eq:sourden2}). The local density can not be much smaller than $10^{-5}$ Mpc${}^{-3}$ as can be estimated from the absence of ``repeaters'' in CR data~\cite{Waxman:1996hp,Kashti:2008bw}. Moreover, the distant to the closest source can not be much larger than, say, 100~Mpc since UHE CRs will unlikely survive over longer distances. 

%%%%%%%%%%%%%%%%%
\begin{table}[t]\centering
\renewcommand{\arraystretch}{1.2}
\begin{tabular}{ccc}
\hline\hline
$D_\star$ [Mpc]&$\mathcal{H}_0$ [$10^{-5}$ Mpc${}^{-3}$]& $L$ [$10^{40}$ erg/s]\\\hline
4&47&2\\8&6&17\\16&0.7&137\\
\hline\hline
\end{tabular}
\caption[]{The local source density $\mathcal{H}_0$ (Eq.~\ref{eq:H0}) and the source luminosity above $10^{19}$~eV (Eq.~\ref{eq:L19}) assuming a comoving distance $D_\star$ to the closest CR source.}\label{tab1}
\end{table}
%%%%%%%%%%%%%%%%%

A spatially homogenous distribution of CR sources with number density $\mathcal{H}(z)$ per comoving volume as in Eq.~(\ref{eq:sourden2}) is equivalent to a diffuse flux of
\begin{equation}
J(E) \simeq \frac{1}{4\pi}\int {\rm d}z\frac{{\rm d} \mathcal{V}}{{\rm d} z}\mathcal{H}(z)J^\star(z,E)\,,
\end{equation}
where $\mathcal{V} = 4\pi d^3_C(z)/3$ is the volume of the co-moving sphere containing the sources at redshift smaller than $z$. In the the following we will assume that the closest CR source is located at a comoving distance $D_\star = d_C(z_\star)$. If the local source distribution with density $\mathcal{H}_0$ is sufficiently smooth we can expect that this is the only source within a distance $D_1=d_C(z_1)>D_\star$ given by,
\begin{equation}\label{eq:H0}
\mathcal{H}_0^{-1} \simeq \int_0^{z_1}{\rm d}z\frac{4\pi d^2_C(z)}{H(z)}(1+z)^n \simeq \frac{4\pi}{3}D_1^3\,.
\end{equation}
We can then decompose the diffuse flux as
\begin{equation}\label{eq:PSdiffuse}
J(E) \simeq \frac{1}{4\pi}J^\star(z_\star,E) + \frac{1}{4\pi}\int_{z_1}^{z_{\rm max}} {\rm d}z\frac{{\rm d} \mathcal{V}}{{\rm d} z}\mathcal{H}(z)J^\star(z,E)\,.
\end{equation}
We assume that the integral can be approximated by the diffuse flux of a spatially homogenous emission rate density $\mathcal{L}(z,E) = \mathcal{H}(z)Q_A(E)$ with $z_{\rm min}= z_1$ as discussed in section~\ref{sec:propagation}. 

%%%%%%%%%%%%%%%%%
\begin{figure}[t]\centering
\includegraphics[width=0.9\linewidth]{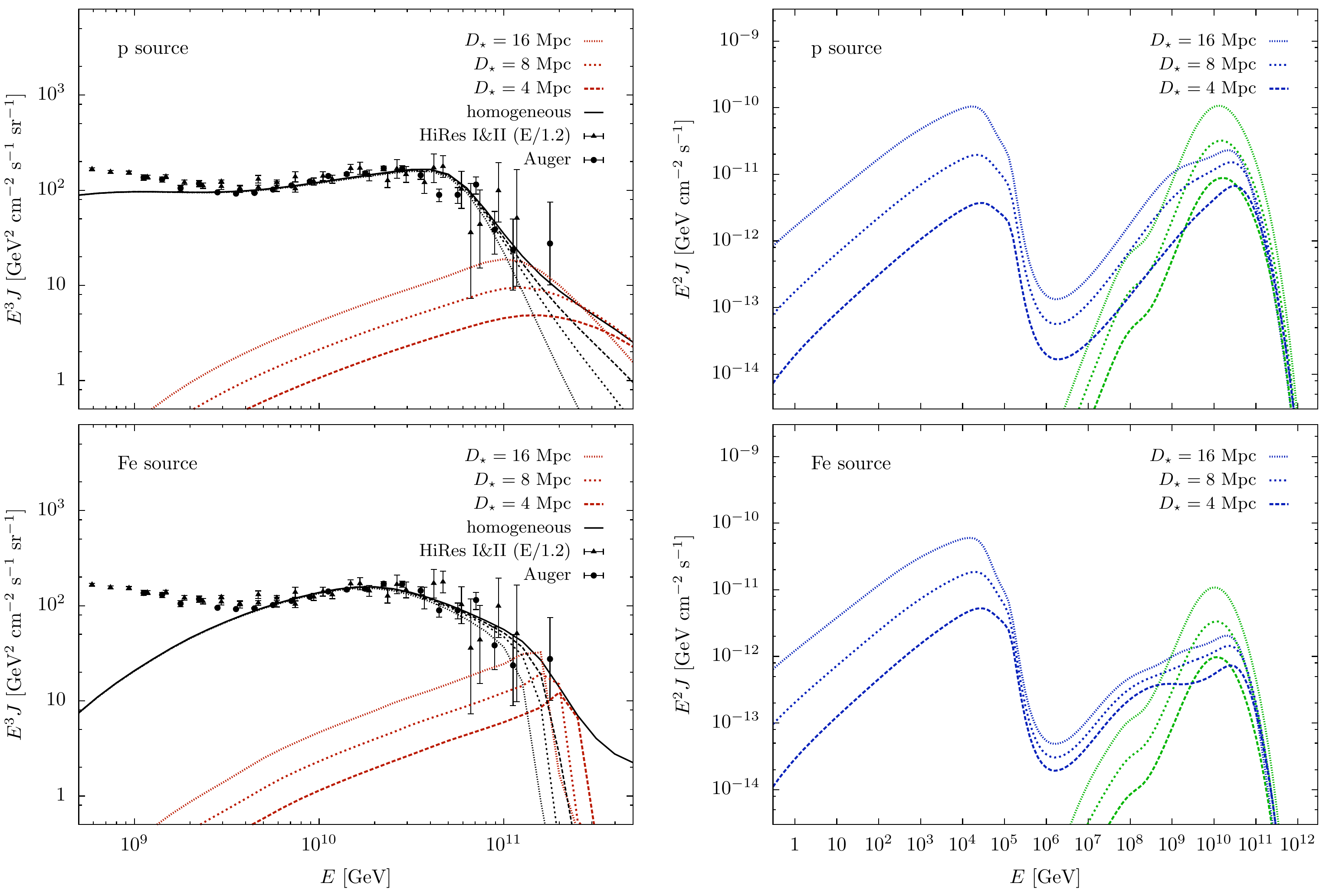}
\caption[]{{\bf Left panels:} The average contribution of the closest proton (top panel) or iron (bottom panel) source to the spectrum of CRs. We assume that the closest source at $z_\star$ determines the average local source density $\mathcal{H}_0$ as in Eq.~\ref{eq:sourden2}. For illustration we assume local densities of $10^{-3}$, $10^{-4}$ and $10^{-3}$~Mpc${}^{-3}$ with $D_\star=$4, 8 and 16~Mpc, respectively. From the fit to the CR data we can determine then the average source luminosity $L$. {\bf Right panels:} The point-source $\gamma$-ray (blue) and neutrino (green) spectra for the closest proton (top panel) or iron (bottom panel) source. We assume again three different distances $D_\star=$4, 8 and 16~Mpc. Whereas the {\it diffuse} $\gamma$-ray spectra shown in Fig.~\ref{fig1} of the two CR models differ by two orders of magnitude, the PS spectra are similar in magnitude for equidistant source locations. The PS fluxes of cosmogenic neutrinos depending on photo-pion interactions of protons are one order of magnitude lower for the case of an all iron source compared to an all proton source.}\label{fig3}
\end{figure}
%%%%%%%%%%%%%%%%

The emission of the closest source at $D_\star$ will only contribute to the UHE CR data at the upper end of the spectrum since $\gamma\ll3$. Due to the very poor statistic at these energies the position can only be determind within large statistical uncertainties. Instead -- as our working hypothesis -- we will assume in the following that the source location is fixed at $D_\star = D_1/2$. This corresponds to the average distance of a source uniformly distributed within a radius $D_1$ and weighted by the flux factor $D^{-2}$. With this choice the sum~(\ref{eq:PSdiffuse}) will closely resemble the CR spectrum from a fully homogenous emission rate density with $z_{\rm min}=0$ as we will see in the following. We will study in the following three different source locations, $D_\star=4$, 8 and 16~Mpc. The corresponding source luminosities and volumes are tabulated in Table~\ref{tab1}.

%%%%%%%%%%%%%%%%%
\begin{figure}[t]\centering
\includegraphics[height=0.35\linewidth,clip=true,viewport=0 0 315 245]{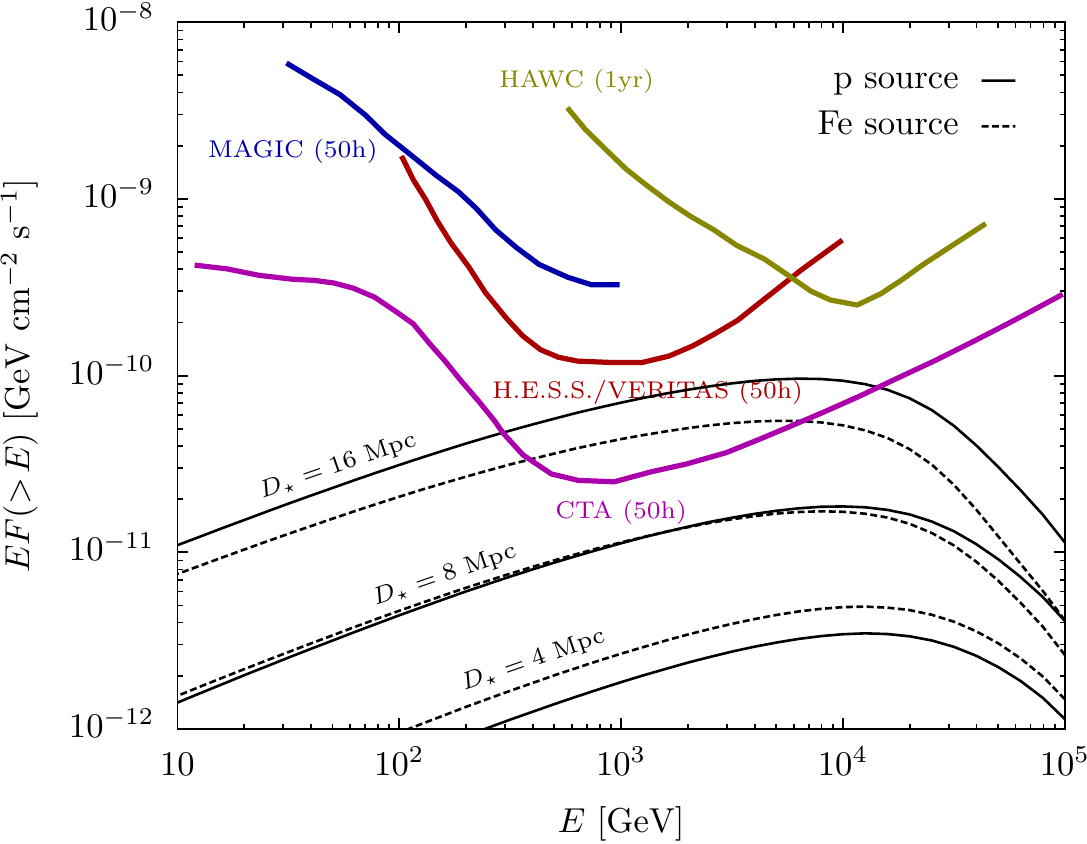}\hspace{0.5cm}\includegraphics[height=0.35\linewidth,clip=true,viewport=0 -2 305 247]{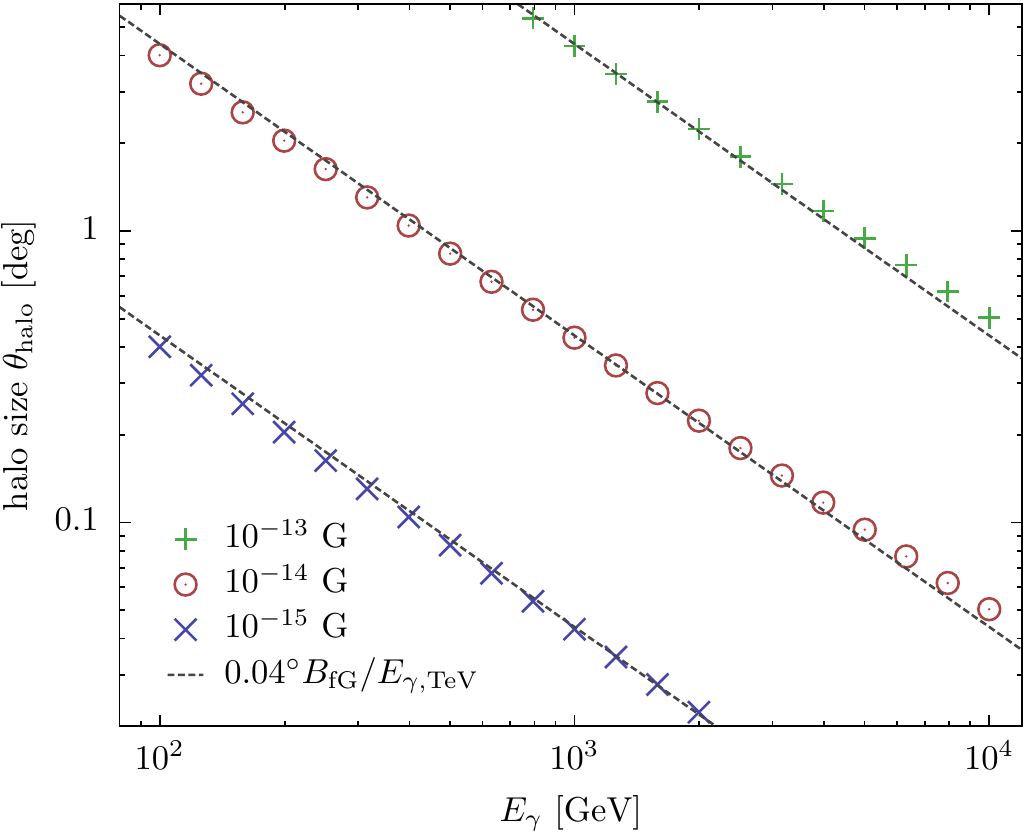}
\caption[]{{\bf Left panel:} The sensitivity of present and future $\gamma$-ray observatories to nearby sources of UHE CRs for the all-proton and all-iron model. The integrated flux $F$ of the point-source is almost independent of the the CR model considered. For a dilute UHE CR source density with ${\mathcal H}_0\simeq10^{-5}$~Mpc${}^{-3}$ the future CTA should be able to identify the cosmogenic $\gamma$-ray flux of UHE CR sources within 50 hours of operation. {\bf Right panel:} The size of the $\gamma$-ray halo for various IGMF strengths and $\gamma$-ray energies. The halo size can be well approximated by a fit $\theta_{\rm halo}\sim0.04^\circ B_{\rm fG}/E_{\gamma, {\rm TeV}}$.}\label{fig4}
\end{figure}
%%%%%%%%%%%%%%%%%

The left panels of Fig.~\ref{fig3} show the contribution of the closest UHE CR sources (red lines) to the overall diffuse flux of UHE CRs for the case of the all-proton (top panel) and the all-iron (bottom panel) model. We also show as thin black lines the remaining contribution of the homogenous CR distribution beyond $D_1$. Note, that in contrast to the PS proton spectra the total PS flux form the iron sources show a strong cutoff preceded by a small bump~\cite{Anchordoqui:1997rn}. This limits the distance to the nearest UHE CR iron source to a few 10~Mpc in this model. The right panels of Fig.~\ref{fig3} show the corresponding PS fluxes in cosmogenic gamma rays (blue lines) and neutrinos (green lines). The PS $\gamma$-ray flux from these close CR sources has the typical $E_\gamma^{-3/2}$ form extending up to several tens of TeV following from inverse Compton emission of a fully Comptonized electron spectrum ($E_e^{-2}$). 

The left panel of Fig.~\ref{fig4} shows the integrated $\gamma$-ray flux of these nearby CR sources in comparison with the sensitivity of present imaging atmospheric Cherenkov telescopes (IACTs) H.E.S.S.~\cite{deOnaWilhelmi:2009zz}, MAGIC~\cite{LopezMoya:2010zz} and VERITAS~\cite{Weekes:2010zz}, as well as the estimated future sensitivity of the water Cherenkov telescope HAWC~\cite{Sinnis:2010zz} and the Cherenkov Telescope Array (CTA)~\cite{Hermann:2010zz}. The solid and dashed lines show the $\gamma$-ray flux from proton and iron sources, respectively, at various distances. The signal depends only weakly on the compostion of the source. More important is the increased luminosity of the source ($\propto D_\star^3$) in the scenario of an increase local source density ($\propto D_\star^{-3}$). For our three CR scenarios shown in Table~\ref{tab1} only a source distribution with a small local source density close to $\mathcal{H}_0 \simeq 10^{-5}$~Mpc${}^{-3}$ ($D_\star=16$~Mpc) and hence a large associated CR luminosity of $L\simeq 10^{42}$~erg/s per source is expected to be visible in the future CTA after 50h of observation. 

Note, however, that our ansatz $D_\star = D_1/2$ has been chosen for a good reproduction of the spatially homogenous emission density shown as the black lines in Fig.~\ref{fig3}. It does not account for stochastic effects of the nearby source distribution. In general, the $\gamma$-ray signal of the closest CR source at comoving distance $D_\star$ and with fixed luminosity $L$ is proportional to $L/D_\star$ and hence the model lines shown in Fig.~\ref{fig4} are expected to shift accordingly. Also, the detection of these multi-TeV $\gamma$-ray point sources in IACTs requires that the signal remains ``point-like'', {\it i.e.}~within the point-spread function (PSF) of the telescope. In this case it is important to consider the effect of an IGMFs on the development of the cascade as we will do in the next section.

\section{Effect of the Intergalactic Magnetic Field}\label{sec:magneticfields}

The cosmogenic $\gamma$-ray cascade of nearby CR sources can only contribute to a GeV-TeV PS flux if the deflections of secondary $e^\pm$ in the cascade via an IGMF is sufficiently small. We can estimate the extend of the cascaded $\gamma$-ray emission by simple geometric arguments following~\cite{Neronov:2007zz}. Deflection of electrons and positrons will be small if the energy loss length $\lambda_e$ of $e^\pm$ via inverse Compton scattering (ICS) is much smaller than the Larmor radius given as $R_L = E/eB \simeq{1.1} (E_{\rm TeV}/B_{\rm fG}){\rm Mpc}$. (Here and in the following we use the abbreviations $E = E_{\rm TeV}{\rm TeV}$, etc.) For center of mass energies much lower than the electron mass, corresponding to energies below PeV in the CMB frame, electrons and positrons interact quickly on kpc scales but with low inelasticity proportional to their energy, $\lambda_e\simeq 0.4~{\rm Mpc}/E_{\rm TeV}$. The typical size of deflections of electrons and positrons is hence $\theta\sim\lambda_e/R_L \sim 0.2^\circ B_{\rm fG}/E^2_{\rm 10TeV}$.

Deflection of $e^\pm$ close to the source have a smaller effect on the size of the halo then deflections close to the observer. To first order, if the cascade experiences a deflection $\Delta \theta$ at a distance $r$ from the observer, we can approximate the corresponding angular displacement $\Delta \theta'$ in the observer's frame via $\Delta \theta'/\Delta\theta \simeq (d-r)/d$. We can account for this scaling in the cascade equation by introducing the corresponding scaling in the Larmor radius $R_L' \simeq R_L d/(d-r)$ or, equivalently, by a scaling of the diffusion matrix of the form ${\mathcal D}' \simeq ((d-r)/d)^2 {\mathcal D}$ (see Appendix~\ref{sec:appI}). Since the distance of the closest CR source is expected to be smaller than the energy loss length by BH pair production or by photo-pion production the electrons and positrons will be produced continuously between the source and the observer. Hence, the average geometric suppression factor of the deflections is $\sim 2$ and hence $\langle R_L'\rangle \simeq 2 R_L$.

In the GeV-TeV energy region the size of the $\gamma$-ray halo is almost independent of the source composition. This is a result of the rapid energy loss of $e^\pm$ via inverse Compton scattering above a few TeV compared to the slow production rate via BH loss of CRs or via pair production of $\gamma$-rays. The leptons quickly lose energy via ICS with CMB photons at a rate $b_{\rm ICS} = E/\lambda_e$; their spectrum in quasi-equilibrium ($\partial_tY_e\simeq 0$) follows the differential equation $\partial_E(b_{\rm ICS}Y_e) \simeq \Gamma_{\rm PP}Y_\gamma$. Thus, the Comptonized electron spectrum for $E\ll E_{\rm max}$ has the form $Y_e\sim E_e^{-2}$. The deflection of an electron of the Comptonized spectrum is approximately $\theta_e \sim \lambda_{e}/R_L/8$ following from $\partial_t(\theta_e Y_e) \simeq Y_e/R_L$ and $Y_e\sim E_e^{-2}$ and $\langle R_L'\rangle \simeq 2 R_L$. The typical photon energy from ICS of a background photon with energy $\epsilon$ is given by $E_\gamma \simeq \epsilon(E_e/m_e)^2$ and hence the halo is expected to extend up to an angle of about $0.01^\circ B_{\rm fG}\epsilon_{\rm meV}/E_{\gamma, {\rm TeV}}$ -- independent of CR composition and source distance.

We can define the halo size more rigorously with the approach outlined in Ref.~\cite{Ahlers:2011jt}. We find that the halo size can be well approximated by the first moment of the angular distribution as
\begin{equation}\label{eq:halo}
\theta_{\rm halo} \equiv \sqrt{2Y_\gamma^{(1)}/Y_\gamma^{(0)}}\,.
\end{equation}
The result of the diffusion-cascade equation (see Appendix~\ref{sec:appI}) is shown in the right panel of Fig.~\ref{fig4}. A fit to the data gives a numerical value of $\theta_{\rm halo} \simeq 0.04^\circ B_{\rm fG}/E_{\gamma, {\rm TeV}}$. This is consistent with our previous estimate for the most abundant meV photons in the CMB spectrum. The typical size of the point-spread function (PSF) of IACTs is of the order of $\theta_{\rm PSF}\simeq0.1^\circ$. Hence, an IGMF with a strength less than $10^{-14}$~G will not significantly decrease the sensitivity of future IACTs to the multi-TeV cosmogenic $\gamma$-ray signal of nearby CR sources.

On the other hand, the non-observation of cascaded $\gamma$-rays as a GeV-TeV PS flux could imply a lower limit on the IGMF strength~\cite{Aharonian:1993vz}. The observation of this effect requires that the sub-TeV $\gamma$-ray emission of the source is relatively quite, such that the cascaded spectrum would dominate the primary flux. Recently, the absence of (resolvable) GeV emission from TeV $\gamma$-ray blazars has been used to infer a lower limits on the IGMF strength at the level $10^{-15}$G ~\cite{Neronov:1900zz,Tavecchio:2010mk}. 

\begin{figure}[t]\centering
\includegraphics[width=0.9\linewidth]{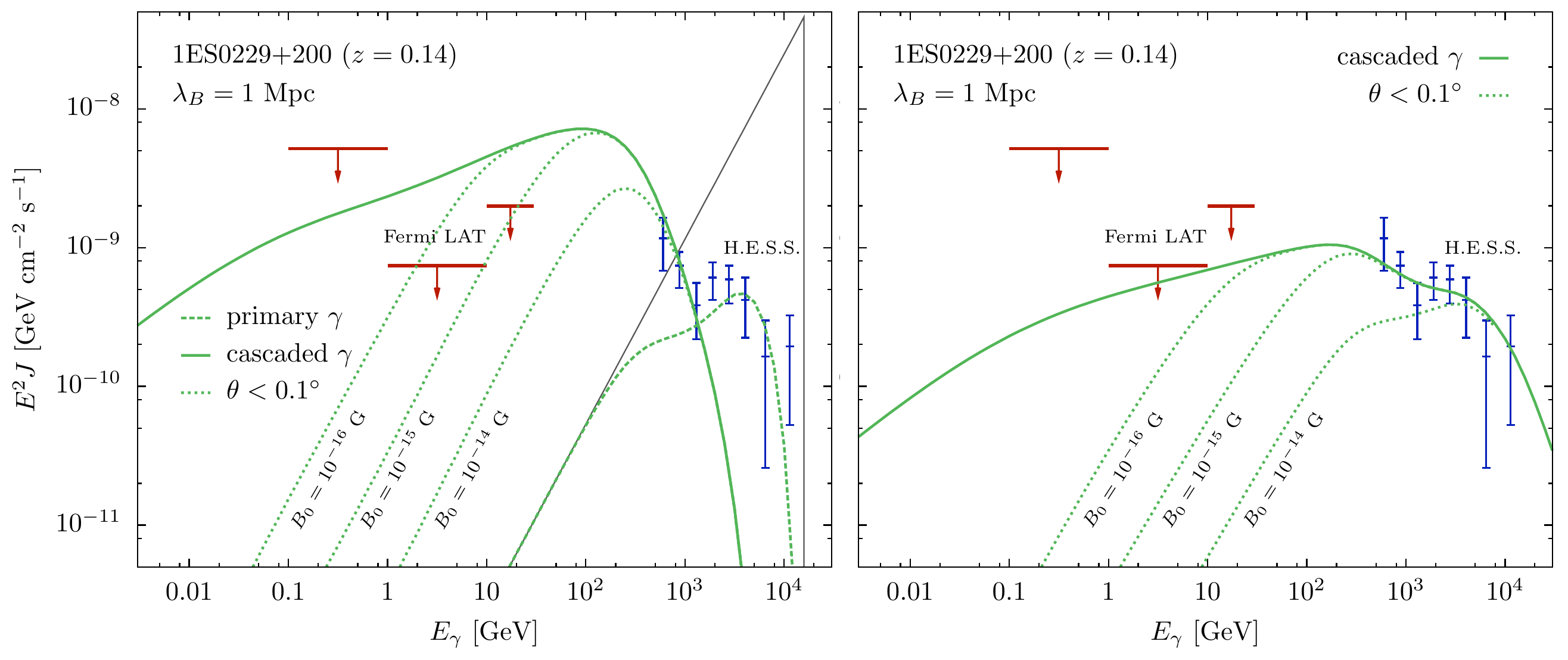}
\caption[]{Two models for the $\gamma$-ray spectra of the blazar source 1ES0229+200 located at redshift $z=0.14$. The green data points show the H.E.S.S.~observation and the green lines the estimated upper flux limits from the non-observation by Fermi-LAT inferred by Ref.~\cite{Tavecchio:2010mk}. {\bf Left panel:} A model for the $\gamma$-ray spectrum assuming $\gamma$-ray emission at a rate $Q_\gamma \propto E^{-2/3}\Theta(20{\rm TeV}-E)$. The solid green line shows the spectrum of secondary $\gamma$-rays without deflections in the IGMF. The dotted green lines indicate the part of the cascaded $\gamma$-ray spectrum within $0.1^\circ$ around the source for an IGMF with coherence length $\lambda_B=1$~Mpc and strength $B_0=10^{-16}$~G, $10^{-15}$~G and $10^{-14}$~G, respectively.  {\bf Right panel:} As in the left panel but now showing the $\gamma$-ray contribution from electromagnetic cascades assuming that the blazar is a CR proton source. We assume that the primary $\gamma$-ray emission is negligible and the (beamed) luminosity in protons is $L_{{\rm p}}\simeq10^{45}$~erg/s. In this case the $\gamma$-ray flux is already below the Fermi-LAT upper limits and no IGMF is required to explain the data.}\label{fig5}
\end{figure}

As an example, we consider here the emission of the blazar source 1ES0229+200 located at redshift $z=0.14$, which has been detected by its TeV $\gamma$-ray emission by H.E.S.S.~\cite{Aharonian:2007wc}. The spectrum is shown in Fig.~\ref{fig5} as the blue data. We consider two models for the $\gamma$-ray observation. In the right panel of Fig.~\ref{fig5} (from Ref.~\cite{Ahlers:2011jt}) we show a model assuming $\gamma$-ray emission from the source with a rate $Q_\gamma \propto E^{-2/3}\Theta(20{\rm TeV}-E)$ (thin gray line). The surviving primary $\gamma$-rays are shown as a dashed green line and secondary cascaded $\gamma$-rays by a solid line. The cascaded spectrum would clearly dominate the sub-TeV emission and is inconsistent with upper limits from Fermi LAT (from Ref.~\cite{Tavecchio:2010mk}). However, the signal within the PSF is significantly reduced by the presence of an IGMF as indicated by the dotted lines. 

%%%%%%%%%%%%%%%%%
%%%%%%%%%%%%%%%%% 

The right panel of Fig.~\ref{fig5} shows an alternative model for the $\gamma$-ray emission of the blazar assuming strong emission of CR protons as in Eq.~(\ref{eq:QA}) with $\gamma=2.3$ and $E_{\rm max}=10^{20.5}$~eV. This model has been advocated in Refs.~\cite{Essey:2009zg,Essey:2009ju,Essey:2010er}. In this case the observed spectrum is assumed to be dominated by cosmogenic $\gamma$-rays emitted during propagation. However, it is apparent that this model does not necessarily require the presence of an IGMF to be compatible with the Fermi-LAT limit as already noted by Ref.~\cite{Essey:2010er}. We also show in this case the reduction of the PS signal via the presence of an IGMF with coherence length $\lambda_B=1$~Mpc and a strength $B_0=10^{-16}$~G, $10^{-15}$~G and $10^{-14}$~G, respectively. 

Note, that the required luminosity of the source is high in this case, $L_{\rm p}\sim10^{45}$~erg/s, which is at least three orders of magnitude larger than the average luminosity of UHE CR sources inferred from the fit to the CR spectrum assuming a local source density larger than $10^{-5}$~Mpc${}^{-3}$. In this model the blazar 1ES0229+200 can hence not be a typical source of UHE CRs. The CR emission along the blazar jet with opening angle $\delta$ and the increase of the effective luminosity as $2/(1-\cos\delta)$ does not play a role in this consideration since the same effect will also decrease the effective local source density of the anisotropically emitting CR sources. However, this example illustrates the strong model dependence on lower limits on the IGMF strength inferred by this method.

\section{Conclusion}\label{sec:conclusion}

We have discussed the production of cosmogenic $\gamma$-rays in models of UHE CR nuclei. These $\gamma$-rays are a result of electromagnetic cascades in the CRB initiated by photo-pion production and BH pair production of the CR nuclei. The signal has its strongest contribution in the GeV to TeV range and is independent of CR interactions in the source environment prior to emission. We have discussed in detail how the $\gamma$-ray flux depend on injection spectra and the chemical composition of the sources. In general, we find that the flux of cosmogenic $\gamma$-rays is less model dependent than other agents of the CR interactions like cosmogenic neutrinos.

As an illustration, we have studied two CR models of the UHE CR spectrum: an all-proton model with strong cosmic evolution and low transition to galactic CRs and an all-iron model dominating the spectrum beyond the ankle. The diffuse $\gamma$-ray flux from these models differs by two orders of magnitude. Whereas the proton model saturates the diffuse extragalactic $\gamma$-ray spectrum inferred by Fermi-LAT the iron model is practically unobservable in the background. We have shown that this large difference in the energy density of the cascade relies on the strong contribution of distant sources assumed in the all-proton model.

The closest sources of UHE CRs can be observed via their $\gamma$-ray point source flux if the IGMF is sufficiently weak ($B_0\lesssim10^{-14}$G). We have argued that the $\gamma$-ray signal is expected to show only small variations with respect to the CR emission model due to the strong contribution of BH pair production. The absolute $\gamma$-ray flux depends on the CR luminosity and position of the source which can be related to the fit to the CR data and by estimates of the (average) local source density. We have estimated that the closest CR source should be observable via its cosmogenic $\gamma$-ray emission in the future Cherenkov Telescope Array if the local source density is small (${\mathcal H}_0\sim10^{-5}$~Mpc${}^{-3}$) and hence the average source luminosity sufficiently large ($L\sim10^{42}$~erg/s). 

We have also briefly commented on the possibility that the TeV emission of distant blazars can be naturally explained as a cosmogenic $\gamma$-ray signal if the blazar is a strong CR proton source ($L\sim10^{45}$~erg/s). As an example we have studied the GeV-TeV emission of the blazar 1ES0229+200. The absence of strong GeV emission of this source has been used to derive lower limits on the IGMF strength. In contrast, the cosmogenic $\gamma$-ray emission of this source is consistent with the observation without the presence of an IGMF.

We have only considered in this study steady sources of CRs, {\it i.e.}~sources which have constant emission during the time of observation. Pulsed sources of UHE CRs could have a stronger $\gamma$-ray emission during the time of activity that could exceed our estimates. The sources of CRs are also expected to emit TeV $\gamma$-rays by CR interactions in the source environment. This contribution may dominate the point-source flux making the observation of cosmogenic $\gamma$-rays difficult. However, in the case of pulsed CR sources, deflections of the cascade via magnetic fields can lead to a time-delay of cosmogenic $\gamma$-rays with respect to the {\it in situ} $\gamma$-ray emission of the source and may help to disentangle the contributions.

\section*{Acknowledgments}

We would like to thank L.~A.~Anchordoqui and M.~C.~Gonzalez-Garcia for valuable comments on the manuscript. MA~would like to thank the {\it Departament d'Estructura i Constituents de la Mat\`eria and Institut de Ciencies del Cosmos} at the University of Barcelona for its hospitality during stages of this project. JS would like to thank the {\it C.N.~Yang Institute for Theoretical Physics}, SUNY, Stony Brook for its hospitality. We acknowledge upported by US NSF Grant No PHY-0969739 and Spanish MICINN grants 2007-66665-C02-01, consolider-ingenio 2010 grant CSD-2008-0037 and CUR Generalitat de Catalunya grant 2009SGR502.

\appendix

\section{Diffusion-Cascade Equations}\label{sec:appI}

The small magnetic deflection of electrons and positrons in weak IGMF with coherence length $\lambda_B$ and strength $B_0$ can be treated as a diffusion process. We follow Ref.~\cite{Ahlers:2011jt} and define the moments of the $\gamma$-ray halo as
\begin{equation}\label{eq:twodimThetan}
{Y}_{e/\gamma}^{(n)} \equiv \frac{2\pi}{(2^nn!)^2}\int\limits_{0}^\infty{\rm d}\theta\,\theta\,\theta^{2n}\,{\mathcal Y}_{e/\gamma}\,.
\end{equation}
One can show that these moments follos the evolution equation
\begin{equation}\label{eq:Thetanevol}
\dot {Y}^{(n)}_\alpha(E) = \partial_E(HE{Y}^{(n)}_\alpha)- \Gamma_\alpha{Y}_{\alpha}^{(n)}(E)+\sum_{\beta=e,\gamma}\int_E{\rm d}E'\gamma_{\beta\alpha}(E',E){Y}_{\beta}^{(n)}(E')
+ \delta_{e\alpha}\int_E{\rm d}E'{\mathcal D}(E',E){Y}_\alpha^{(n-1)}(E')\,,
\end{equation}
with diffusion matrix
\begin{equation}\label{eq:Drandom}
{\mathcal D}(E',E) \simeq \frac{1}{3}\frac{\min(1,\lambda_B\Gamma_{\rm ICS}(E))}{E\,\Gamma_{\rm ICS}(E)}\frac{e^2B_0^2}{E'^2\langle x\rangle(E')}\,.
\end{equation}
The 	quantity $\langle x\rangle$ denotes the inelasticity of of ICS and hence $\langle x\rangle\Gamma_{\rm ICS} = 1/\lambda_{\rm ICS}$. The first moment ${Y}_{e/\gamma}^{(0)}$ equals the PS flux ${J}^\star_{e/\gamma}$ as a solution of the Boltzmann equations~(\ref{eq:boltzmann}).

We define discrete values ${Y}^{(n)}_{e,i} \simeq \Delta E_i {Y}^{(n)}_{e}(E_i)$, $Q_{e,i} \simeq \Delta E_iQ_e(E_i)$, etc. The combined effect of transitions and deflections within the cascade during a sufficiently small time-step $\Delta t$ can be described by the matrix equations
\begin{align}\label{eq:ThetaCas0}
\begin{pmatrix}{Y}_\gamma({t}+\Delta{t})\\{Y}_e({t}+\Delta{t})\end{pmatrix}^{(0)}_i
  &\simeq \sum_j\begin{pmatrix}T_{\gamma\gamma}(\Delta
  t)&T_{e\gamma}(\Delta t)\\T_{\gamma e}(\Delta
  t)&T_{ee}(\Delta
  t)\end{pmatrix}_{ji}\begin{pmatrix}{Y}_\gamma({t})\\{Y}_e({t})\end{pmatrix}^{(0)}_j + \Delta t\begin{pmatrix}Q_\gamma\\Q_e\end{pmatrix}_i\,,\\\label{eq:ThetaCasn}
\begin{pmatrix}{Y}_\gamma({t}+\Delta{t})\\{Y}_e({t}+\Delta{t})\end{pmatrix}^{(n)}_i
  &\simeq \sum_j\begin{pmatrix}T_{\gamma\gamma}(\Delta
  t)&T_{e\gamma}(\Delta t)\\T_{\gamma e}(\Delta
  t)&T_{ee}(\Delta
  t)\end{pmatrix}_{ji}\begin{pmatrix}{Y}_\gamma({t})\\{Y}_e({t})\end{pmatrix}^{(n)}_j + \Delta t\begin{pmatrix}0&0\\0&{\mathcal D}\end{pmatrix}_{ji}\begin{pmatrix}{Y}_\gamma({t})\\{Y}_e({t})\end{pmatrix}^{(n-1)}_j\quad(n>0)\,,
  \end{align}
The full cascade solution is then given by
 \begin{equation}
\begin{pmatrix}{Y}_\gamma({t}')\\{Y}_e({t}')\end{pmatrix}^{(n)}_i
  \simeq 
  \sum\limits_{m=0}^n\sum_j\mathcal{A}^{(m)}_{ji}(t'-t)\begin{pmatrix}{Y}_\gamma({t})\\{Y}_e({t})\end{pmatrix}^{(n-m)}_j + \Delta t\sum_j\mathcal{B}^{(n)}_{ji}(t'-t)\begin{pmatrix}Q_\gamma\\Q_e\end{pmatrix}_j\,.
\end{equation}
The $2n$ matrizes $\mathcal{A}^{(m)}$ and $\mathcal{B}^{(m)}$ follow the recursive relation 
\begin{align}\label{eq:rec1}
\mathcal{A}^{(n)}(2^p \Delta t) &= \sum\limits_{i=0}^n\mathcal{A}^{(i)}(2^{p-1}\Delta t)\cdot\mathcal{A}^{(n-i)}(2^{p-1}\Delta t)\,,\\
\mathcal{B}^{(n)}(2^p\Delta t) &= \mathcal{B}^{(n)}(2^{p-1}\Delta t) +  \sum\limits_{i=0}^n\mathcal{A}^{(i)}(2^{p-1}\Delta t)\cdot\mathcal{B}^{(n-i)}(2^{p-1}\Delta t)\,,\label{eq:rec2}
\end{align}
where the non-zero initial conditions are $\mathcal{A}^{(0)}(\Delta t) = \mathcal{T}(\Delta t)$, $\mathcal{A}^{(1)}_{ij} = {\rm diag}(0,\Delta t \mathcal{D}_{ij})$ and $\mathcal{B}^{(0)}(\Delta t) =\mathbf{1}$. The matrices $\mathcal{A}^{(0)}$ and $\mathcal{B}^{(0)}$ are the familiar transfer matrices for electromagnetic cascades in the presence of a source term. Using the recursion relations (\ref{eq:rec1}) and (\ref{eq:rec2}) we can efficiently calculate the matrices $\mathcal{A}^{(n)}$ and $\mathcal{B}^{(n)}$ via matrix-doubling~\cite{Protheroe:1992dx}.


\begin{thebibliography}{99}

\bibitem{Nagano:2000ve}
  M.~Nagano and A.~A.~Watson,
  %``Observations And Implications Of The Ultrahigh-Energy Cosmic Rays,''
  Rev.\ Mod.\ Phys.\  {\bf 72}, 689 (2000).
  %%CITATION = RMPHA,72,689;%%

\bibitem{Amsler:2008zzb}
  C.~Amsler {\it et al.}  [Particle Data Group],
  %``Review of particle physics,''
  Phys.\ Lett.\  B {\bf 667}, 1 (2008).
  %%CITATION = PHLTA,B667,1;%%  
  
\bibitem{Abraham:2010mj}
  J.~Abraham {\it et al.}  [Pierre Auger Collaboration],
  %``Measurement Of The Energy Spectrum Of Cosmic Rays Above $10^{18}$ Ev Using
  %The Pierre Auger Observatory,''
  Phys.\ Lett.\  B {\bf 685}, 239 (2010)
  [arXiv:1002.1975 [astro-ph.HE]].
  %%CITATION = PHLTA,B685,239;%%

\bibitem{Abraham:2010yv}
  J.~Abraham {\it et al.}  [Pierre Auger Collaboration],
  %``Measurement of the Depth of Maximum of Extensive Air Showers above 10^18
  %eV,''
  Phys.\ Rev.\ Lett.\  {\bf 104}, 091101 (2010)
  [arXiv:1002.0699 [astro-ph.HE]].
  %%CITATION = PRLTA,104,091101;%%
    
\bibitem{Abbasi:2007sv}
  R.~Abbasi {\it et al.}  [HiRes Collaboration],
  %``Observation of the GZK cutoff by the HiRes experiment,''
  Phys.\ Rev.\ Lett.\  {\bf 100}, 101101 (2008)
  [arXiv:astro-ph/0703099].
  %%CITATION = PRLTA,100,101101;%%
  
\bibitem{Abbasi:2009nf}
  R.~U.~Abbasi {\it et al.}  [HiRes Collaboration],
  %``Indications of Proton-Dominated Cosmic Ray Composition above 1.6 EeV,''
  Phys.\ Rev.\ Lett.\  {\bf 104}, 161101 (2010)
  [arXiv:0910.4184 [astro-ph.HE]].
  %%CITATION = PRLTA,104,161101;%%
  
\bibitem{Linsley:1963bk}
  J.~Linsley, {\it Proceedings of ICRC 1963, Jaipur, India}, pp.~77-99
  %``Primary cosmic rays of energy 10**17 to 10**20-eV: The energy spectrum and
  %arrival directions,''

\bibitem{Hill:1983mk}
  C.~T.~Hill and D.~N.~Schramm,
  %``The Ultrahigh-Energy Cosmic Ray Spectrum,''
  Phys.\ Rev.\  D {\bf 31}, 564 (1985).
  %%CITATION = PHRVA,D31,564;%%

\bibitem{Wibig:2004ye}
  T.~Wibig and A.~W.~Wolfendale,
  %``At what particle energy do extragalactic cosmic rays start to
  %predominate?,''
  J.\ Phys.\ G {\bf 31}, 255 (2005).
  [arXiv:astro-ph/0410624].
  %%CITATION = JPHGB,G31,255;%%
  
\bibitem{Berezinsky:2002nc}
  V.~Berezinsky, A.~Z.~Gazizov and S.~I.~Grigorieva,
  %``On astrophysical solution to ultra high energy cosmic rays,''
  Phys.\ Rev.\  D {\bf 74}, 043005 (2006).
  [arXiv:hep-ph/0204357].
  %%CITATION = PHRVA,D74,043005;%%
  
\bibitem{Fodor:2003ph} 
  Z.~Fodor, S.~D.~Katz, A.~Ringwald and H.~Tu,
  %``Bounds on the cosmogenic neutrino flux,''
  JCAP {\bf 0311}, 015 (2003)
  [arXiv:hep-ph/0309171].
  %%CITATION = HEP-PH 0309171;%%
  
\bibitem{Greisen:1966jv}
  K.~Greisen,
  %``End To The Cosmic Ray Spectrum?,''
  Phys.\ Rev.\ Lett.\  {\bf 16}, 748 (1966).
  %%CITATION = PRLTA,16,748;%%

\bibitem{Zatsepin:1966jv}
  G.~T.~Zatsepin and V.~A.~Kuz'min,
  %``Upper limit of the spectrum of cosmic rays,''
  JETP Lett.\  {\bf 4}, 78 (1966)
  [Pisma Zh.\ Eksp.\ Teor.\ Fiz.\  {\bf 4}, 114 (1966)].
  %%CITATION = ZFPRA,4,114;%%

\bibitem{Abraham:2008ru}
  J.~Abraham {\it et al.}  [Pierre Auger Collaboration],
  %``Observation of the suppression of the flux of cosmic rays above $4\times
  %10^{19}$eV,''
  Phys.\ Rev.\ Lett.\  {\bf 101}, 061101 (2008)
  [arXiv:0806.4302 [astro-ph]].
  %%CITATION = PRLTA,101,061101;%%
  
\bibitem{Stecker:1969fw}
  F.~W.~Stecker,
  %``Photodisintegration of ultrahigh-energy cosmic rays by the universal
  %radiation field,''
  Phys.\ Rev.\  {\bf 180}, 1264 (1969).
  %%CITATION = PHRVA,180,1264;%%

\bibitem{Puget:1976nz}
  J.~L.~Puget, F.~W.~Stecker and J.~H.~Bredekamp,
  %``Photonuclear Interactions Of Ultrahigh-Energy Cosmic Rays And Their
  %Astrophysical Consequences,''
  Astrophys.\ J.\  {\bf 205}, 638 (1976).
  %%CITATION = ASJOA,205,638;%%
   
\bibitem{Stecker:1998ib}
  F.~W.~Stecker and M.~H.~Salamon,
  %``Photodisintegration of ultrahigh energy cosmic rays: A new
  %determination,''
  Astrophys.\ J.\  {\bf 512}, 521 (1999)
  [arXiv:astro-ph/9808110].
  %%CITATION = ASJOA,512,521;%%
  
\bibitem{Stecker:1978ah}
  F.~W.~Stecker,
  %``Diffuse Fluxes Of Cosmic High-Energy Neutrinos,''
  Astrophys.\ J.\  {\bf 228}, 919 (1979).
  %%CITATION = ASJOA,228,919;%%
  
\bibitem{Blumenthal:1970nn}
  G.~R.~Blumenthal,
  %``Energy loss of high-energy cosmic rays in pair-producing collisions with
  %ambient photons,''
  Phys.\ Rev.\  D {\bf 1}, 1596 (1970).
  %%CITATION = PHRVA,D1,1596;%%

\bibitem{Anchordoqui:2006pd}
  L.~A.~Anchordoqui, J.~F.~Beacom, H.~Goldberg, S.~Palomares-Ruiz and T.~J.~Weiler,
  %``TeV gamma-rays from photo-disintegration / de-excitation of cosmic-ray
  %nuclei,''
  Phys.\ Rev.\ Lett.\  {\bf 98} (2007) 121101
  [arXiv:astro-ph/0611580].
  %%CITATION = PRLTA,98,121101;%%

\bibitem{Aharonian:2010te}
  F.~Aharonian and A.~M.~Taylor,
  %``Limitations on the Photo-disintegration Process as a Source of VHE
  %Photons,''
  Astropart.\ Phys.\  {\bf 34}, 258 (2010)
  [arXiv:1005.3230 [astro-ph.HE]].
  %%CITATION = APHYE,34,258;%%
  
\bibitem{Franceschini:2008tp}
  A.~Franceschini, G.~Rodighiero and M.~Vaccari,
  %``The extragalactic optical-infrared background radiations, their time
  %evolution and the cosmic photon-photon opacity,''
  Astron.\ Astrophys.\  {\bf 487}, 837 (2008)
  [arXiv:0805.1841 [astro-ph]].
  %%CITATION = ARXIV:0805.1841;%%
 
\bibitem{Ahlers:2009rf}
  M.~Ahlers, L.~A.~Anchordoqui and S.~Sarkar,
  %``Neutrino diagnostics of ultra-high energy cosmic ray protons,''
  Phys.\ Rev.\  D {\bf 79}, 083009 (2009)
  [arXiv:0902.3993 [astro-ph.HE]].
  %%CITATION = PHRVA,D79,083009;%%
  
\bibitem{Goriely:2008zu}
  S.~Goriely, S.~Hilaire and A.~J.~Koning,
  %``Improved predictions of nuclear reaction rates with the TALYS reaction code
  %for astrophysical applications,''
  Astron.\ Astrophys.\  {\bf 487}, 767 (2008)
  [arXiv:0806.2239 [astro-ph]], \url{http://www.talys.eu/}
  %%CITATION = AAEJA,487,767;%%
  
\bibitem{RachenTHESIS}
  J.~P.~Rachen, {\it Interaction processes and statistical properties of the propagation of cosmic-rays in photon backgrounds}, PhD thesis of the Bonn University, 1996. 

\bibitem{Mucke:1999yb}
  A.~M\"ucke, R.~Engel, J.~P.~Rachen, R.~J.~Protheroe and T.~Stanev,
  %``Monte Carlo simulations of photohadronic processes in astrophysics,''
  Comput.\ Phys.\ Commun.\  {\bf 124}, 290 (2000) 
  [arXiv:astro-ph/9903478].
  
\bibitem{Ahlers:2010ty}
  M.~Ahlers and A.~M.~Taylor,
  %``Analytic Solutions of Ultra-High Energy Cosmic Ray Nuclei Revisited,''
  Phys.\ Rev.\  D {\bf 82}, 123005 (2010)
  [arXiv:1010.3019 [astro-ph.HE]].
  %%CITATION = PHRVA,D82,123005;%%

\bibitem{Allard:2008gj}
  D.~Allard, N.~G.~Busca, G.~Decerprit, A.~V.~Olinto and E.~Parizot,
  %``Implications of the cosmic ray spectrum for the mass composition at the
  %highest energies,''
  JCAP {\bf 0810}, 033 (2008)
  [arXiv:0805.4779 [astro-ph]].
  %%CITATION = JCAPA,0810,033;%%
  
\bibitem{Lee:1996fp}
  S.~Lee,
  %``On The Propagation Of Extragalactic High-Energy Cosmic And Gamma-Rays,''
  Phys.\ Rev.\  D {\bf 58}, 043004 (1998)
  [arXiv:astro-ph/9604098].
  %%CITATION = PHRVA,D58,043004;%%
  
\bibitem{Blumenthal:1970gc}
  G.~R.~Blumenthal and R.~J.~Gould,
  %``Bremsstrahlung, synchrotron radiation, and compton scattering of
  %high-energy electrons traversing dilute gases,''
  Rev.\ Mod.\ Phys.\  {\bf 42}, 237 (1970).
  %%CITATION = RMPHA,42,237;%%
    
\bibitem{Protheroe:1996si}
  R.~J.~Protheroe and P.~L.~Biermann,
  %``A new estimate of the extragalactic radio background and implications
  %for ultra-high-energy gamma ray propagation,''
  Astropart.\ Phys.\  {\bf 6}, 45 (1996)
  [Erratum-ibid.\  {\bf 7}, 181 (1996)],
  [arXiv:astro-ph/9605119].
  %%CITATION = ASTRO-PH 9605119;%%

\bibitem{Ahlers:2010fw}
  M.~Ahlers, L.~A.~Anchordoqui, M.~C.~Gonzalez-Garcia, F.~Halzen and S.~Sarkar,
  %``GZK Neutrinos after the Fermi-LAT Diffuse Photon Flux Measurement,''
  Astropart.\ Phys.\  {\bf 34}, 106 (2010)
  [arXiv:1005.2620 [astro-ph.HE]].
  %%CITATION = APHYE,34,106;%%
  
\bibitem{Kronberg:1993vk}
  P.~P.~Kronberg,
  %``Extragalactic magnetic fields,''
  Rept.\ Prog.\ Phys.\  {\bf 57}, 325 (1994).
  %%CITATION = RPPHA,57,325;%%
  
\bibitem{Beck:2008ty}
  R.~Beck,
  %``Galactic and Extragalactic Magnetic Fields,''
  AIP Conf.\ Proc.\  {\bf 1085}, 83 (2009)
  [arXiv:0810.2923 [astro-ph]].
  %%CITATION = APCPC,1085,83;%%
  
\bibitem{Aharonian:1993vz}
  F.~A.~Aharonian, P.~S.~Coppi and H.~J.~Volk,
  %``Very high-energy gamma-rays from AGN: Cascading on the cosmic background
  %radiation fields and the formation of pair halos,''
  Astrophys.\ J.\  {\bf 423}, L5 (1994)
  [arXiv:astro-ph/9312045].
  %%CITATION = ASJOA,423,L5;%%
  
\bibitem{Neronov:2007zz}
  A.~Neronov and D.~V.~Semikoz,
  %``A method of measurement of extragalactic magnetic fields by TeV gamma ray
  %telescopes,''
  JETP Lett.\  {\bf 85}, 473 (2007)
  [arXiv:astro-ph/0604607].
  %%CITATION = JTPLA,85,473;%%
  
\bibitem{d'Avezac:2007sg}
  P.~d'Avezac, G.~Dubus and B.~Giebels,
  %``Cascading on extragalactic background light,''
  Astron.\ Astrophys.\  {\bf 469}, 857 (2007)
  [arXiv:0704.3910 [astro-ph]].
  %%CITATION = AAEJA,469,857;%%
    
\bibitem{Neronov:1900zz}
  A.~Neronov and I.~Vovk,
  %``Evidence for strong extragalactic magnetic fields from Fermi observations
  %of TeV blazars,''
  Science {\bf 328}, 73 (2010).
  [arXiv:1006.3504 [astro-ph.HE]].
  %%CITATION = SCIEA,328,73;%%
  
\bibitem{Tavecchio:2010mk}
  F.~Tavecchio, G.~Ghisellini, L.~Foschini, G.~Bonnoli, G.~Ghirlanda and P.~Coppi,
  %``The intergalactic magnetic field constrained by Fermi/LAT observations of
  %the TeV blazar 1ES 0229+200,''
  Mon.\ Not.\ Roy.\ Astron.\ Soc.\  {\bf 406}, L70 (2010)
  [arXiv:1004.1329 [astro-ph.CO]].
  %%CITATION = MNRAA,406,L70;%%

\bibitem{Kalashev:2007sn}
  O.~E.~Kalashev, D.~V.~Semikoz and G.~Sigl,
  %``Ultra-High Energy Cosmic Rays and the GeV-TeV Diffuse Gamma-Ray Flux,''
  Phys.\ Rev.\  D {\bf 79}, 063005 (2009)
  [arXiv:0704.2463 [astro-ph]].
  %%CITATION = PHRVA,D79,063005;%%
    
\bibitem{Loeb:2000na}
  A.~Loeb and E.~Waxman,
  %``Gamma-Ray Background from Structure Formation in the Intergalactic
  %Medium,''
  Nature {\bf 405}, 156 (2000)
  [arXiv:astro-ph/0003447].
  %%CITATION = NATUA,405,156;%%
  
\bibitem{Abdo:2010dk}
  A.~A.~Abdo {\it et al.}  [Fermi-LAT Collaboration],
  %``Constraints on Cosmological Dark Matter Annihilation from the Fermi-LAT
  %Isotropic Diffuse Gamma-Ray Measurement,''
  JCAP {\bf 1004}, 014 (2010)
  [arXiv:1002.4415 [astro-ph.CO]].
  %%CITATION = JCAPA,1004,014;%%
  
\bibitem{Dermer:2007fg}
  C.~D.~Dermer,
  %``The Extragalactic Gamma Ray Background,''
  AIP Conf.\ Proc.\  {\bf 921}, 122 (2007)
  [arXiv:0704.2888 [astro-ph]].
  %%CITATION = APCPC,921,122;%%

\bibitem{Abdo:2010nz}
  A.~A.~Abdo {\it et al.}  [Fermi-LAT Collaboration],
  %``The Spectrum of the Isotropic Diffuse Gamma-Ray Emission Derived From
  %First-Year Fermi Large Area Telescope Data,''
  Phys.\ Rev.\ Lett.\  {\bf 104}, 101101 (2010)
  [arXiv:1002.3603 [astro-ph.HE]].
  %%CITATION = PRLTA,104,101101;%%

\bibitem{Sreekumar:1997un}
  P.~Sreekumar {\it et al.}  [EGRET Collaboration],
  %``EGRET observations of the extragalactic gamma ray emission,''
  Astrophys.\ J.\  {\bf 494}, 523 (1998)
  [arXiv:astro-ph/9709257].
  %%CITATION = ASJOA,494,523;%%
  
\bibitem{Berezinsky:2010xa}
  V.~Berezinsky, A.~Gazizov, M.~Kachelriess and S.~Ostapchenko,
  %``Restricting UHECRs and cosmogenic neutrinos with Fermi-LAT,''
  Phys.\ Lett.\  B {\bf 695}, 13 (2011)
  [arXiv:1003.1496 [astro-ph.HE]].
  %%CITATION = PHLTA,B695,13;%%

\bibitem{Abbasi:2011ji}
  R.~Abbasi {\it et al.}  [IceCube Collaboration],
  %``Constraints on the Extremely-high Energy Cosmic Neutrino Flux with the
  %IceCube 2008-2009 Data,''
  [arXiv:1103.4250 [astro-ph.CO]].
  %%CITATION = ARXIV:1103.4250;%%
  
\bibitem{Armengaud:2005cr}
  E.~Armengaud, G.~Sigl and F.~Miniati,
  %``Gamma Ray Astronomy with Magnetized Zevatrons,''
  [arXiv:astro-ph/0511277].
  %%CITATION = ASTRO-PH/0511277;%%
  
\bibitem{Kotera:2010xd}
  K.~Kotera, D.~Allard and M.~Lemoine,
  %``Detectability of ultrahigh energy cosmic ray signatures in gamma rays,''
  Astron.\ Astrophys.\  {\bf 527}, A54+ (2011)
  [arXiv:1011.0575 [astro-ph.HE]].
  %%CITATION = ARXIV:1011.0575;%%
  
\bibitem{Ferrigno:2004am}
  C.~Ferrigno, P.~Blasi and D.~De Marco,
  %``High energy gamma ray counterparts of astrophysical sources of  ultrahigh
  %energy cosmic rays,''
  Astropart.\ Phys.\  {\bf 23}, 211 (2005)
  [arXiv:astro-ph/0404352].
  %%CITATION = APHYE,23,211;%%

\bibitem{Gabici:2005gd}
  S.~Gabici and F.~A.~Aharonian,
  %``Point-like gamma ray sources as signatures of distant accelerators of ultra
  %high energy cosmic rays,''
  Phys.\ Rev.\ Lett.\  {\bf 95}, 251102 (2005)
  [arXiv:astro-ph/0505462].
  %%CITATION = PRLTA,95,251102;%%
  
\bibitem{Essey:2009zg}
  W.~Essey and A.~Kusenko,
  %``A new interpretation of the gamma-ray observations of active galactic
  %nuclei,''
  Astropart.\ Phys.\  {\bf 33}, 81 (2010)
  [arXiv:0905.1162 [astro-ph.HE]].
  %%CITATION = APHYE,33,81;%%

\bibitem{Essey:2009ju}
  W.~Essey, O.~E.~Kalashev, A.~Kusenko and J.~F.~Beacom,
  %``Secondary photons and neutrinos from cosmic rays produced by distant
  %blazars,''
  Phys.\ Rev.\ Lett.\  {\bf 104}, 141102 (2010)
  [arXiv:0912.3976 [astro-ph.HE]].
  %%CITATION = PRLTA,104,141102;%%

\bibitem{Essey:2010er}
  W.~Essey, O.~Kalashev, A.~Kusenko and J.~F.~Beacom,
  %``Role of line-of-sight cosmic ray interactions in forming the spectra of
  %distant blazars in TeV gamma rays and high-energy neutrinos,''
  Astrophys.\ J.\  {\bf 731}, 51 (2011)
  [arXiv:1011.6340 [astro-ph.HE]].
  %%CITATION = ASJOA,731,51;%%

\bibitem{Gelmini:2007jy}
  G.~BGelmini, O.~E.~Kalashev, D.~V.~Semikoz,
  %``GZK Photons Above 10-EeV,''
  JCAP {\bf 0711}, 002 (2007).
  [arXiv:0706.2181 [astro-ph]].
  
\bibitem{Hooper:2010ze}
  D.~Hooper, A.~M.~Taylor, S.~Sarkar,
  %``Cosmogenic photons as a test of ultra-high energy cosmic ray composition,''
  Astropart.\ Phys.\  {\bf 34}, 340-343 (2011).
  [arXiv:1007.1306 [astro-ph.HE]].
  
\bibitem{Waxman:1996hp}
  E.~Waxman, K.~B.~Fisher and T.~Piran,
  %``The signature of a correlation between > 10**19-eV cosmic ray sources  and
  %large scale structure,''
  Astrophys.\ J.\  {\bf 483}, 1 (1997)
  [arXiv:astro-ph/9604005].
  %%CITATION = ASJOA,483,1;%%
  
\bibitem{Kashti:2008bw}
  T.~Kashti and E.~Waxman,
  %``Searching for a Correlation Between Cosmic-Ray Sources Above 10^{19} eV and
  %Large-Scale Structure,''
  JCAP {\bf 0805}, 006 (2008)
  [arXiv:0801.4516 [astro-ph]].
  %%CITATION = JCAPA,0805,006;%%

\bibitem{Anchordoqui:1997rn}
  L.~A.~Anchordoqui, M.~T.~Dova, L.~N.~Epele and J.~D.~Swain,
  %``A depression before the bump in the highest energy cosmic ray spectrum,''
  Phys.\ Rev.\  D {\bf 57}, 7103 (1998)
  [arXiv:astro-ph/9708082].
  %%CITATION = PHRVA,D57,7103;%%
  
\bibitem{deOnaWilhelmi:2009zz}
  E.~de Ona Wilhelmi  [HESS Collaboration],
  %``Status of the H.E.S.S. telescope,''
  AIP Conf.\ Proc.\  {\bf 1112}, 16 (2009).
  %%CITATION = APCPC,1112,16;%%

\bibitem{LopezMoya:2010zz}
  M.~Lopez Moya  [MAGIC Collaboration],
  %``Scientific highlights and status of the MAGIC telescope,''
  AIP Conf.\ Proc.\  {\bf 1223}, 99 (2010).
  %%CITATION = APCPC,1223,99;%%
  
\bibitem{Weekes:2010zz}
  T.~C.~Weekes {\it et al.}  [VERITAS Collaboration],
  %``VERITAS: Status Summary 2009,''
  Int.\ J.\ Mod.\ Phys.\  D {\bf 19}, 1003 (2010).
  %%CITATION = IMPAE,D19,1003;%%
  
\bibitem{Sinnis:2010zz}
  G.~Sinnis  [HAWC and Milagro Collaborations],
  %``Water Cherenkov Technology In Gamma-Ray Astrophysics,''
  Nucl.\ Instrum.\ Meth.\  A {\bf 623}, 410 (2010).
  %%CITATION = NUIMA,A623,410;%%
  
\bibitem{Hermann:2010zz}
  G.~Hermann  [CTA Collaboration],
  %``Towards The Future Cherenkov Telescope Array Cta,''
  Nucl.\ Instrum.\ Meth.\  A {\bf 623}, 408 (2010).
  %%CITATION = NUIMA,A623,408;%%
  
\bibitem{Ahlers:2011jt}
  M.~Ahlers,
  %``Gamma-ray halos as a measure of intergalactic magnetic fields: a classical
  %moment problem,''
  [arXiv:1104.5172 [astro-ph.HE]].
  %%CITATION = ARXIV:1104.5172;%%
  
\bibitem{Aharonian:2007wc}
  F.~Aharonian {\it et al.}  [HESS Collaboration],
  %``New constraints on the Mid-IR EBL from the HESS discovery of VHE gamma rays
  %from 1ES 0229+200,''
  Astron.\ Astrophys.\  {\bf 475},L9-L13 (2007)
  [arXiv:0709.4584 [astro-ph]].
  %%CITATION = ARXIV:0709.4584;%%
  
\bibitem{Protheroe:1992dx}
  R.~J.~Protheroe and T.~Stanev,
  %``Electron photon cascading of very high energy gamma-rays in the  infrared
  %background,''
  Mon. Not. R. Astron. Soc. {\bf 264}, 191 (1993).
  
\end{thebibliography}
\end{document}